\documentclass[submission, Phys]{SciPost}

\hypersetup{
    colorlinks,
    linkcolor={red!50!black},
    citecolor={blue!50!black},
    urlcolor={blue!80!black}
}

\usepackage[bitstream-charter]{mathdesign}
\DeclareSymbolFont{usualmathcal}{OMS}{cmsy}{m}{n}
\DeclareSymbolFontAlphabet{\mathcal}{usualmathcal}

\usepackage{caption}
\usepackage{subcaption}
\newcommand{\saja}{\textsc{SaJa}}
\newcommand{\klf}{\textsc{KLFitter}}
\newcommand{\ttbar}{$t\bar{t}$}
\newcommand{\madgraph}{\textsc{MG5\_aMC@NLO}}
\newcommand{\powheg}{\textsc{POWHEG}}
\newcommand{\pythia}{\textsc{Pythia8}}
\newcommand{\herwig}{\textsc{Herwig++}}

\begin{document}

\begin{center}{\Large \textbf{
Zero-Permutation Jet-Parton Assignment using a Self-Attention Network\\
}}\end{center}

\begin{center}
Jason S. H. Lee\textsuperscript{1},
Inkyu Park\textsuperscript{1},
Ian J. Watson\textsuperscript{1*} and
Seungjin Yang\textsuperscript{1}
\end{center}

\begin{center}
{\bf 1} Department of Physics, University of Seoul, Seoul 02504, Republic of Korea
\\
${}^\star$ {\small \sf ijwatson@physics.uos.ac.kr}
\end{center}

\begin{center}
\today
\end{center}


\section*{Abstract}
{\bf
In high-energy particle physics events, it can be advantageous to find the jets associated with the decays of intermediate states, for example, the three jets produced by the hadronic decay of the top quark.
Typically, a goodness-of-association measure, such as a $\chi^2$ related to the mass of the associated jets, is constructed, and the best jet combination is found by optimizing this measure.
As this process suffers from a combinatorial explosion with the number of jets, the number of permutations is limited by using only the $n$ highest $p_T$ jets.
The self-attention block is a neural network unit used for the neural machine translation problem, which can highlight relationships between any number of inputs in a single iteration without permutations.
In this paper, we introduce the \href{https://github.com/CPLUOS/SaJa}{Self-Attention for Jet Assignment (\saja) network}.
\saja\ can take any number of jets for input and outputs probabilities of jet-parton assignment for all jets in a single step.
We apply \saja\ to find jet-parton assignments of fully-hadronic \ttbar\ events to evaluate the performance.
We show that \saja\ achieves better performance than a likelihood-based approach.
}

\vspace{10pt}
\noindent\rule{\textwidth}{1pt}
\tableofcontents\thispagestyle{fancy}
\noindent\rule{\textwidth}{1pt}
\vspace{10pt}

\section{Introduction}
\label{sec:intro}
  The final state of proton--proton collisions is an explosion of particles.
  To tame the complexity of the description of the collision, we can order the particles by clustering them into jets, final state collections of particles that are produced from an underlying parton in the hard interaction.
  In events that contain heavy intermediate particles, such as the top quark, which can decay into three partons, it can be useful to find the jets correlated with the decay of the underlying particle.
  Typically, a goodness-of-association measure, such as a $\chi^2$ or likelihood related to the mass of the associated jets, is constructed, and jet combinations are trialed in turn to find the combination with the optimal measure value which is used as the assignment for further analysis~ \cite{sirunyan2019measurement,aad2015measurement,erdmann2014likelihood}.
  This process suffers from a combinatorial explosion as the number of jets can get quite high and usually only combinations of the $n$ highest $p_T$ jets are tried, where $n$ is usually restricted to at most six or seven.
  Recently, machine learning algorithms such as boosted decision trees and fully-connected neural networks have also been used to learn a goodness-of-association measure for jet-parton assignment~\cite{erdmann2017jet,erdmann2019bottom}.
  However, while these approaches show better performance than existing kinematic fitting methods, they still suffer from the problem of combinatorial explosion as they also operate on individual permutations of the jets.

  Recent advances in deep neural networks have led us to a solution that analogizes the jet-parton association problem to finding connections between words in a sentence, and how words in the sentence are translated to another language~\cite{arXiv:1409.0473,arXiv:1606.01933,arXiv:1703.03906,arXiv:1706.03762}.
  As languages do not map one-to-one, words corresponding to different concepts can appear at different points in the sentence and its translation.
  In our case, the source language corresponds to the reconstructed jets and the target language is the parton assignment for each jet.
  The self-attention block is a neural network unit used in the machine translation problem, which can highlight relationships between any number of inputs in a single step. 
  
  In this paper, we introduce the Self-Attention for Jet Assignment (\saja) network.
  As the name implies, it uses self-attention blocks to find the relationships between jets in the event.
  \saja\ can take any number of jets for input and outputs the probability of association to the possible jet-parton assignment categories for all jets.
In section~\ref{sec:method} we describe the \saja\ network and the procedure we use to optimize the network.
  The fully hadronically decaying $t\bar{t}$ channel is one of the standard model processes which produces the most number of jets, which makes the traditional permutation-based methods limited due to the combinatorics.
  In section~\ref{sec:analysis} we describe the datasets we use to study this process and the analysis procedure.
  Finally, in section~\ref{sec:results}, we show the results of applying \saja\ to the fully hadronic $t\bar{t}$ channel and compare the performance with a likelihood-based approach.

\section{The \saja\ Network}
\label{sec:method}

This section describes the implementation of \saja.
We first explain the objective function that will be used to optimize the \saja\ network.
Then, the self-attention block at the heart of \saja\ is described.
Finally, the architecture of \saja\ as a whole is shown.
The code we used to implement \saja\ has also been made publicly available~\cite{seungjin_yang_2020_4311381}.

\subsection{The Objective Function}
\label{sec:objective}

We consider the jet-parton assignment problem as a jet-wise multi-class classification task.
That is, we will build a model which, for each jet in the event, gives a probability for the jet to belong to each of several classes.
For our example of fully hadronically decaying top pair production, there are five classes: $b$ jets produced in the decay $t \to bW$, light quark jets from the decay $W \to jj'$, $b$ jets, and $W$ jets produced from the equivalent anti-top decays, and jets not in association with decays of the top quarks, which we will call other jets.
Since it is hard to distinguish jets originating from \(t\) and jets originating from \(\bar{t}\) when both tops are hadronically decaying, we introduce arbitrary indices 1 and 2 for either of the top quarks in the pair and their decay products and develop an objective function which is insensitive to the ordering of the jets.

The model is thus a function $f^\theta$ of the form:
\begin{equation}\label{eq:model_eq}
    f^{\theta}:
    \begin{pmatrix}
        \mathbf{x}^{(1)} \\
        \vdots \\
        \mathbf{x}^{(N)}
    \end{pmatrix}
    \rightarrow
    \begin{pmatrix}
        \hat{y}^{(1)}_{b_{1}} & \hat{y}^{(1)}_{W_{1}} & \hat{y}^{(1)}_{b_{2}} & \hat{y}^{(1)}_{W_{2}} & \hat{y}^{(1)}_{\textrm{other}} \\
        \vdots & \vdots & \vdots & \vdots & \vdots \\
        \hat{y}^{(N)}_{b_{1}} & \hat{y}^{(N)}_{W_{1}} & \hat{y}^{(N)}_{b_{2}} & \hat{y}^{(N)}_{W_{2}} & \hat{y}^{(N)}_{\textrm{other}}
    \end{pmatrix}
\end{equation}
where \(\theta\) represents the trainable model parameters, \(\mathbf{x}^{ \left(j\right) }\) denotes the jet labelled by an index \(j\) in the event, $\hat{y}^{(j)}_{\textrm{class}}$ indicates the probability for the jet \(j\) to belong to the category class, and the corresponding genuine category is represented by a one-hot encoding; each label is denoted by \(y_{class}^{(j)}\).

When using the output of the model for jet assignment, the jet probability is passed through the argmax function to assign each jet a single class.
If the assignments do not match the topology of $t\bar{t}$ production, that is if there is anything \emph{other} than exactly one $b_{1}$, two $W_{1}$, one $b_{2}$, two $W_{2}$ with the remaining of the jets assigned to the other class, the event is rejected. These rejected events are referred to as topologically invalid.

Our task is jet-wise classification, therefore we can use the average of the cross-entropy for all jets in the event as the objective function.
Since our indices are arbitrary, there remains an ambiguity by permuting $t$ and $\bar{t}$ between indices 1 and 2. 
Therefore, we write the objective function $J(\theta)$ as:
\begin{gather}\label{eq:objective}
    J(\theta) = \frac{1}{N} \sum_{j=1}^{N} \left ( \min{(\pi_{12}^{(j)}, \pi_{21}^{(j)})} - y_{ \textrm{other} }^{(j)} \log{\hat{y}_{\textrm{other}}^{(j)}} \right )
\end{gather}
where
\(\pi_{\alpha\beta}^{(j)} = -\left[
y_{b}^{(j)} \log{\hat{y}_{b_{\alpha}}^{(j)}}
+ y_{\bar{b}}^{(j)} \log{\hat{y}_{b_{\beta}}^{(j)}} 
+ y_{W^{+}}^{(j)} \log{\hat{y}_{W_{\alpha}}^{(j)}}
+ y_{W^{-}}^{(j)} \log{\hat{y}_{W_{\beta}}^{(j)}}
\right]\)
with \(\alpha\), \(\beta \in \{1,2\}\).
As the $\min$ function is invariant under the permutation of arguments, Eq.~\ref{eq:objective} is independent of the top or anti-top being assigned to the first or second top.

\subsection{Self-Attention}

Self-attention is a special case of the more general attention mechanism, of which there are several implementations such as additive attention~\cite{arXiv:1409.0473} and dot-product attention~\cite{arXiv:1703.03906}.
In this paper, we use the scaled dot-product attention which was introduced in the Transformer network~\cite{arXiv:1706.03762}.
The scaled dot-product attention is a function that takes three sequences of vectors as inputs and produces a single output sequence.
Since these sequences will be permutation-invariant under the self-attention mechanism, we call the sequences \emph{sets}.
The three input sets of the attention block are called keys, values, and queries, which are notated by $K$, $V$, and $Q$ respectively.
The cardinalities of $K$ and $V$ are required to be the same and will be denoted $S$, while the cardinality of $Q$ may differ and will be denoted $T$.
A set is represented as a matrix, where each row is a vector of the set.
In general, the key set and the value set are derived from a single set called the source, which is the jet inputs in our case.
The source set is transformed into a key and value set using two different element-wise linear transformations.
Self-attention is the special case where the query set is also derived from the source set in the same manner.
In summary, the sets derived from the source set are:
\begin{equation}\label{eq:att-1-KV-same}
    K = \begin{pmatrix}
        \vec{k}^{(1)}\\
        \vdots \\
        \vec{k}^{(S)}
    \end{pmatrix}
    = \begin{pmatrix}
        W_{K}\vec{x}^{(1)}\\
        \vdots \\
        W_{K}\vec{x}^{(S)}
    \end{pmatrix}
    ,\ 
    V = \begin{pmatrix}
        \vec{v}^{(1)}\\
        \vdots \\
        \vec{v}^{(S)}
    \end{pmatrix} = \begin{pmatrix}
        W_{V}\vec{x}^{(1)}\\
        \vdots \\
        W_{V}\vec{x}^{(S)}
    \end{pmatrix} ,\ 
    Q = \begin{pmatrix}
        \vec{q}^{(1)}\\
        \vdots \\
        \vec{q}^{(T)}
    \end{pmatrix} = \begin{pmatrix}
        W_{Q}\vec{x}^{(1)}\\
        \vdots \\
        W_{Q}\vec{x}^{(T)}
    \end{pmatrix}
\end{equation}
where $\vec{x}^{(k)}$ is a vector of the source set, and $W_{K}$, $W_{V}$, $W_{Q}$ are linear transformation matrices.
To form the self-attention layer, first $Q$ and $K$ are matrix multiplied and then scaled by the dimension of a key vector $d_{k}$ to form the matrix $E$.
Then the softmax function is applied to each row of $E$ and the elements of the resulting matrix $A$ are called the attention weights:
\begin{equation}\label{eq:att-3-softmax}
    A_{ij} = \frac{e^{E_{ij}}}{Z_{i}}, \textrm{ where }  
    E_{ij} = \frac{1}{\sqrt{d_{k}}} \vec{q}^{(i)} \cdot \vec{k}^{(j)} \textrm{ and }
    Z_{i} = \sum_{j=1}^{N} e^{E_{ij}}
\end{equation}
If the scaling is not applied in eq.~\ref{eq:att-3-softmax}, the magnitude of $E_{ij}$ increases as $d_{k}$ increases. As $E_{ij}$ increases, the gradient of $A_{ij}$ goes to zero.
Thus, the scaling in eq.~\ref{eq:att-3-softmax} is a simple remedy for the vanishing gradient problem.

Finally, the attention weights are multiplied by the values and the outputs are summed to get the output of the self-attention layer $H_i$:
\begin{equation}\label{eq:att-4-context}
    H_{i} = {\sum_{j=1}^{S} A_{ij} \vec{v}^{(j)}}
\end{equation}
Therefore, the self-attention layer is a weighted sum of the elements of the input set and is insensitive to the ordering of the elements.

In real-world usage, multiple attention functions are computed in parallel to increase the representation power.
The multi-head attention mechanism has multiple branches (called heads), where the above self-attention is performed for each head separately on the input, resulting in multiple $H_i$ outputs.
These outputs are concatenated into a single matrix along the feature axis, then a final element-wise linear transformation is applied, resulting in the output of the multi-head attention layer.

\subsection{Architecture}

\begin{figure}[tbp]
    \centering
    \begin{subfigure}[b]{0.3\textwidth}
        \includegraphics[width=\textwidth]{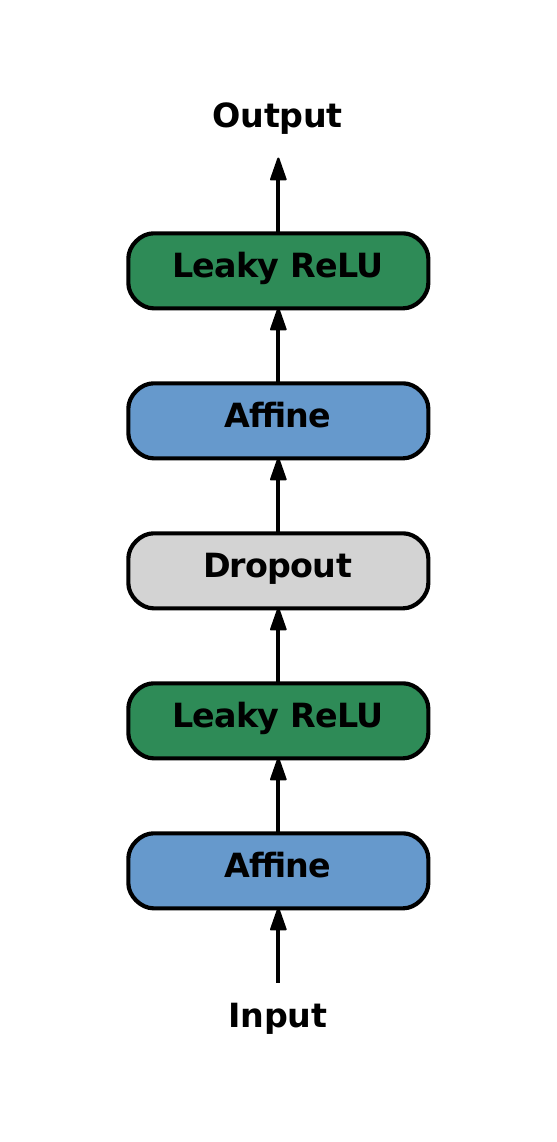}
    \end{subfigure}
    \hfill  
    \begin{subfigure}[b]{0.3\textwidth}
        \includegraphics[width=\textwidth]{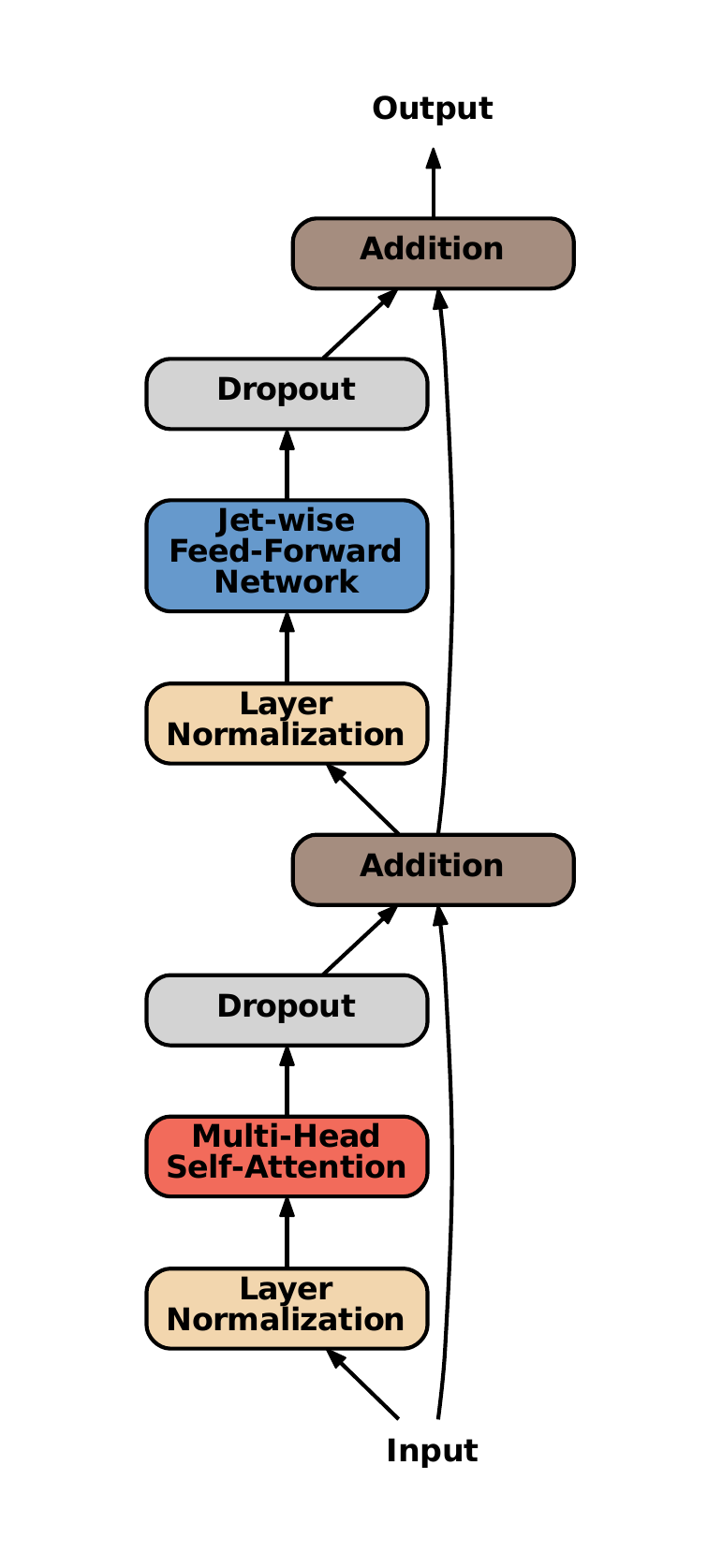}
    \end{subfigure}
    \hfill
    \begin{subfigure}[b]{0.3\textwidth}
        \includegraphics[width=\textwidth]{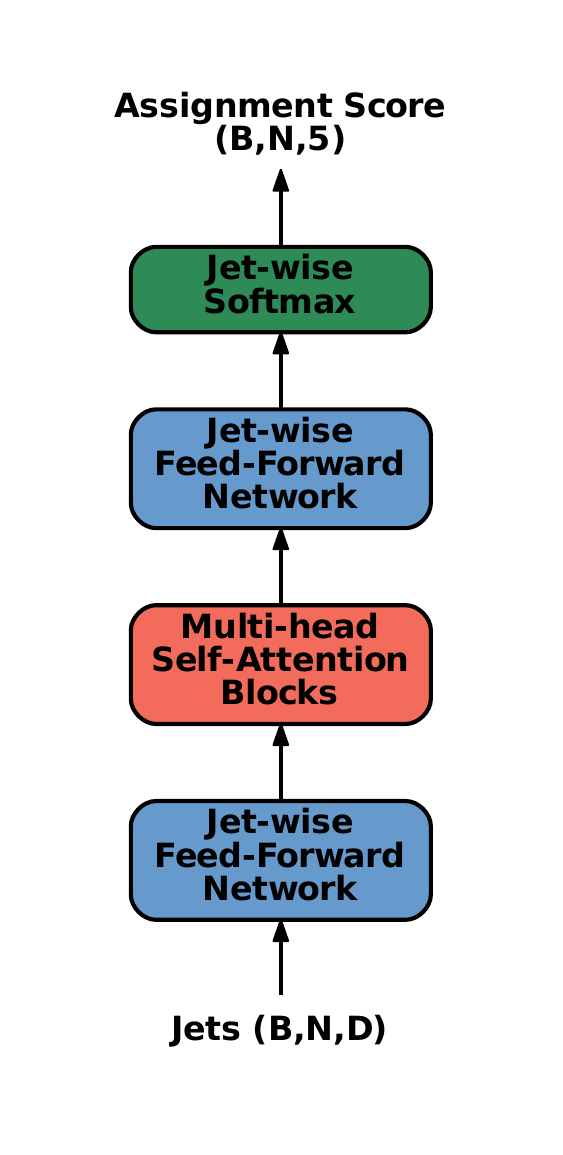}
    \end{subfigure}
  \hfill

  \caption{\label{fig:architecture} The data flows of the jet-wise feed-forward network (left), multi-head self-attention block (center), and \saja\ (right) are shown. \(B\) denotes the size of the batch. \(N\) indicates the maximum number of jets in the batch. \(D\) indicates the number of features representing the jet.}
\end{figure}

The architecture of \saja\ and its two building blocks are shown in figure~\ref{fig:architecture}.
The two building blocks are as follows.
The feed-forward network consists of an affine transformation, followed by a leaky ReLU~\cite{arXiv:1803.08375}, then a dropout~\cite{arXiv:1207.0580}, and another affine transformation followed by a leaky ReLU.
The multi-head self-attention block consists of the multi-head self-attention layer and another feed-forward network, each of which is sandwiched between layer normalization~\cite{arXiv:1607.06450} and dropout layers, and surrounded by a skip connection~\cite{arXiv:1512.03385}.

At the bottom of \saja, a feed-forward layer takes a set of vectors (jets) as the input.
The first affine transformation layer produces vectors with dimension of $d_{\textrm{FFN}}$ and the second one produces vectors with dimension of $d_{\textrm{model}}$. 
The output of the bottom feed-forward network is passed through a stack of $N_{\textrm{block}}$ multi-head self-attention blocks, each of which has $N_{\textrm{head}}$ heads.
A top feed-forward layer followed by jet-wise softmax takes the output of a last multi-head self-attention block and then produces the assignment score.
The dimension of the first affine transformation output is $d_{\textrm{FFN}}$ and the second one is five, which corresponds to the number of classes.
Therefore, with this setup, the hyperparameters of \saja\ are only $d_{\textrm{model}}$, $d_{\textrm{FFN}}$, $N_{\textrm{block}}$, and $N_{\textrm{head}}$, where $d_{\textrm{model}}=d_{k} \times N_{\textrm{head}}$.

All operations, except for the self-attention block, are jet-wise operations.
Since the self-attention block is invariant under jet permutations, \saja\ itself is invariant under jet permutations.
This property can be used to open the black box of \saja\ with Monte Carlo information, by ordering the inputs by the truth labels without affecting the resulting output.
The two main operations of \saja\ are affine transformations and the attention function.
They both consist of matrix multiplications, which can be computed through highly optimized routines and accelerated by GPUs.

\subsection{Predictive uncertainty}

Uncertainty quantification in deep learning is becoming more important as deep learning is applied to the real world~\cite{arXiv:1606.06565}.
Poor predictions by deep learning models can pose danger, especially when deep learning is being deployed in situations such as for self-driving cars and medical diagnoses.
Models can also be presented with out-of-distribution (OOD) data, which is data that does not follow the distribution of the training set data.
In these cases, an uncertainty estimate can be used to override a decision made by the AI.
Similarly, \saja\ can make wrong assignments in fully hadronic \ttbar\ events, resulting in poor resolution of the kinematics of the reconstructed top quark.
\saja\ will also be exposed to background events such as QCD multijet during the inference (physics analysis) phase.
Therefore, we will use deep learning uncertainty to veto both poor jet-parton assignments on fully hadronic \ttbar\ events and background events.

There are various types of uncertainties that have been proposed for deep learning models \cite{gal2016uncertainty}, from which we will use the predictive entropy.
As \saja\ does not make a prediction for a single object, it simultaneously predicts the class of all jets in the event, we estimate an uncertainty on the jet-parton assignment using the average of predictive entropies for all jets, \(H[\hat{Y}]\):
\begin{equation}\label{eq:entropy}
    \mathbb{H}[\hat{Y}] = \frac{1}{N} \sum_{j=1}^{N} \left ( -\sum_{c \in \textrm{classes} } \hat{y}_{c}^{(j)} \log \hat{y}_{c}^{(j)} \right )
\end{equation}
We will use this predictive entropy as an event selection.
When the jet-parton assignment results in a predictive entropy higher than a threshold, which means the network has given an uncertain assignment for all the jets in the event, the event is rejected.

One of the most promising applications of deep learning uncertainty is to detect OOD test data.
In this study, multijet background events exactly correspond to an OOD sample.
In many cases, the prediction score of modern deep learning classifiers is not calibrated and can classify OOD data as an in-distribution class with high confidence~\cite{arXiv:1706.04599}.
However, \saja\ shows reasonable uncertainty distributions for all physics processes without any calibration method, as will be shown in section~\ref{sec:results}.

\section{Monte Carlo Samples and Event Selection}
\label{sec:analysis}


We simulate all-hadronic \ttbar\ pair production with up to two additional jets in the final state from $pp$ collisions at \(\sqrt{s}\) = 13 TeV using \textsc{MadGraph5\_aMC@NLO} v2.2.2 at next to leading order~\cite{Alwall:2011uj}.
In the event generation, the top quark mass is assumed to be 172.5 GeV.
We also generate multijet events using \textsc{MadGraph\_aMC@NLO} at leading order with two to four partons in the final state, which are the main background for the all-hadronic \(t\bar{t}\) event analysis.
The hard process events are interfaced to \textsc{Pythia} 8.212 which simulates the parton shower and hadronization~\cite{Sjostrand:2014zea}.
The hadronized events are then processed by \textsc{Delphes} v3.4.2~\cite{deFavereau:2013fsa}, which performs a fast detector response simulation and particle-flow event reconstruction for a CMS-like detector and uses \textsc{FastJet 3.3.2} to perform anti-\(k_{T}\) jet clustering~ \cite{Cacciari:2011ma}.
We use the default CMS card provided with the \textsc{Delphes} package, except that the jet-clustering parameter \(R\) is modified from 0.5 to 0.4 to match the value that CMS uses for 13~TeV data.

We also generate $t\bar{t}$ events using \textsc{POWHEG}v2~\cite{alioli2010general,Frixione:2007nw,Nason:2004rx,Frixione:2007vw} interfaced to both \textsc{Pythia}8.212 and \textsc{Herwig++} 2.7.1~\cite{bahr2008herwig++} to check the Monte Carlo dependency of our results. To match the matrix element and parton shower, we use $h_{\text{\text{damp}}}$ values of $1.581$ for \textsc{Pythia8}~\cite{CMS-PAS-TOP-16-021} and $1$ for \textsc{Herwig++}.



We follow the trigger selection used in the CMS all-hadronic top pair analysis to select events for further study~\cite{sirunyan2019measurement}.
We use jets with \(p_{T} > 30\) GeV and \(|\eta| < 2.4\).
We select events that contain at least six jets of \( p_{T} \) above 40 GeV, and at least one of the jets is b-tagged.
We then require $H_{T}=\sum_{\text{jets}}p_{T}>450\,\text{GeV}$.

We generated about 19.4 million \ttbar\ events using \textsc{MadGraph5\_aMC@NLO} with \textsc{Pythia8}, of which about 2.4 million events passed the event selection.
We generated multijet events in seven \(H_T\) bins to increase the statistics at high \(H_T\). 
In total, 157.5 million multijet events were generated, of which 1.1 million passed the event selection.


\begin{figure}[tbp]
    \centering
    \includegraphics[width=0.65\textwidth]{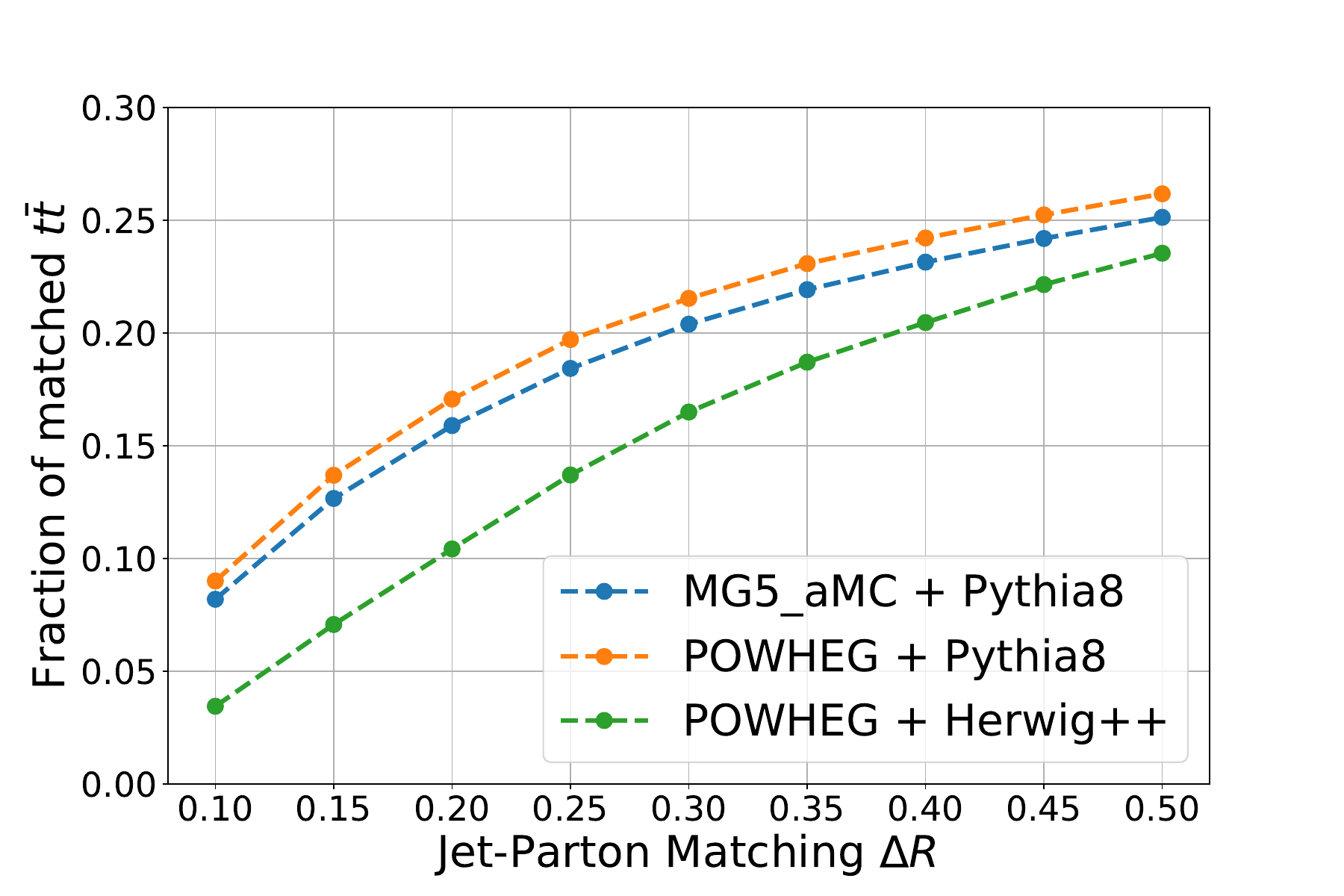}
    \caption{Fraction of matched \ttbar\, events after the event selection vs. angular distance threshold between a parton and a jet $\Delta R(\text{jet},\text{parton})$. \textsc{MadGraph\_aMC@NLO} interfaced to \textsc{Pythia8} (blue), \textsc{POWHEG} interfaced to \textsc{Pythia8} (orange), and \textsc{POWHEG} interfaced to \textsc{Herwig++} (green). \textsc{MadGraph\_aMC@NLO} interfaced to \textsc{Pythia8} with $\Delta R_{\text{max}}=0.3$ events are used for training.}
    \label{fig:jet-parton-match}
\end{figure}

To study the performance of the jet-parton assignment, we match the reconstructed jets to the outgoing partons of the matrix elements.
We perform a geometric matching between jets and partons in \(\eta-\phi\) space requiring matched jets and partons to be separated by less than 0.3 in \(\Delta R\) and by pairing a jet and a parton at the nearest distance first and removing both from the list of potential matches.
Figure~\ref{fig:jet-parton-match} shows the fraction of matched \ttbar\ events as a function of $\Delta R (\text{jet},\text{parton})$ for the three event generators.
\textsc{POWHEG} v2 with \textsc{Pythia8} has the fraction about 1\% higher than \textsc{MadGraph5\_aMC@NLO} with \textsc{Pythia8}.
The difference between \textsc{POWHEG} v2 with \textsc{Herwig++} and \textsc{MadGraph5\_aMC@NLO} with \textsc{Pythia8} decrease from about 5.6\% to 1.6\% as $\Delta R$ increases.
In about 20\% of selected top quark pair events, all partons are matched with jets.
We call top pair events where the jet-parton matching failed to match all the partons unmatched events and consider them as background.

\begin{figure}[tbp]
    \centering
    \includegraphics[width=0.48\textwidth]{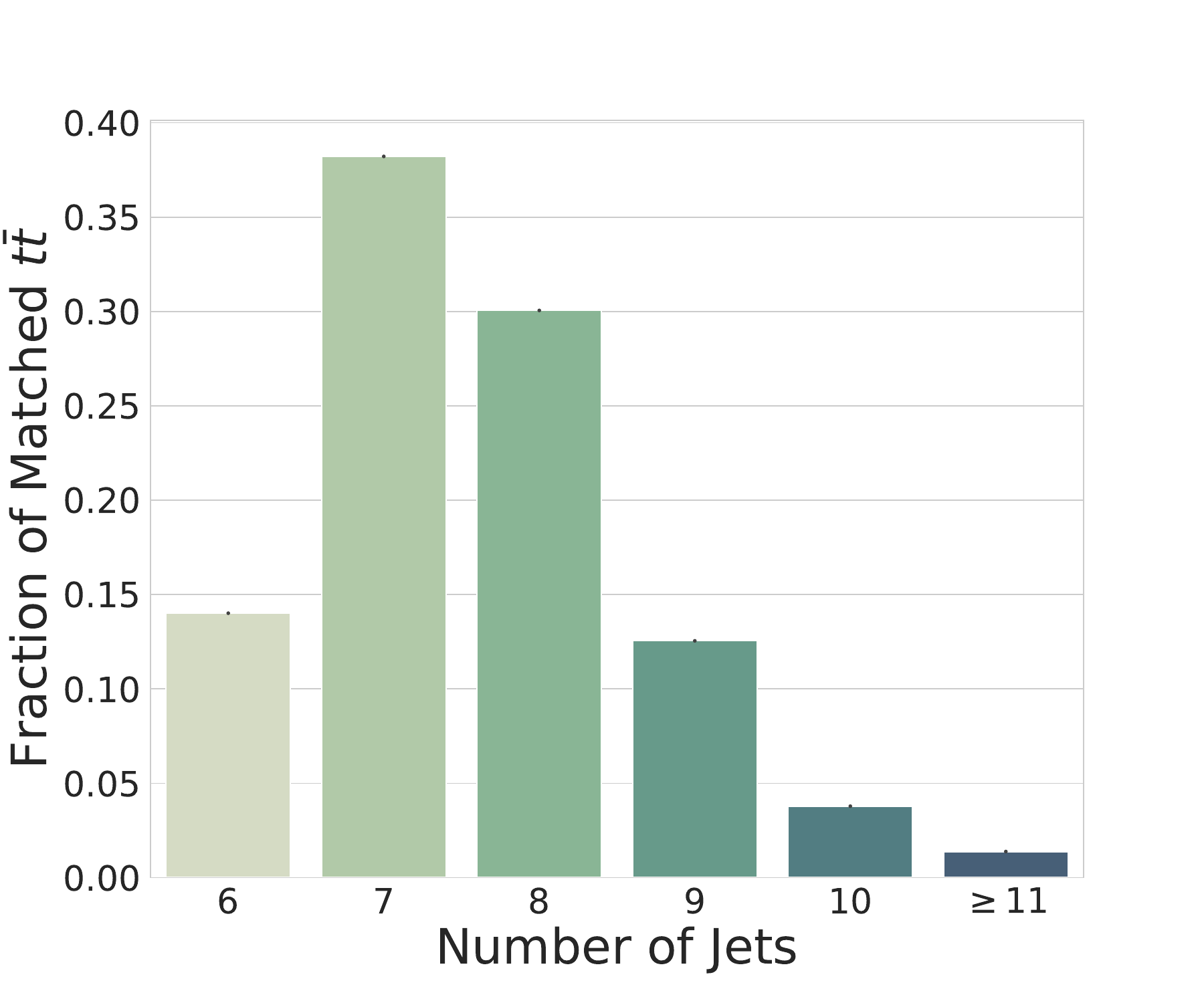}
    \includegraphics[width=0.48\textwidth]{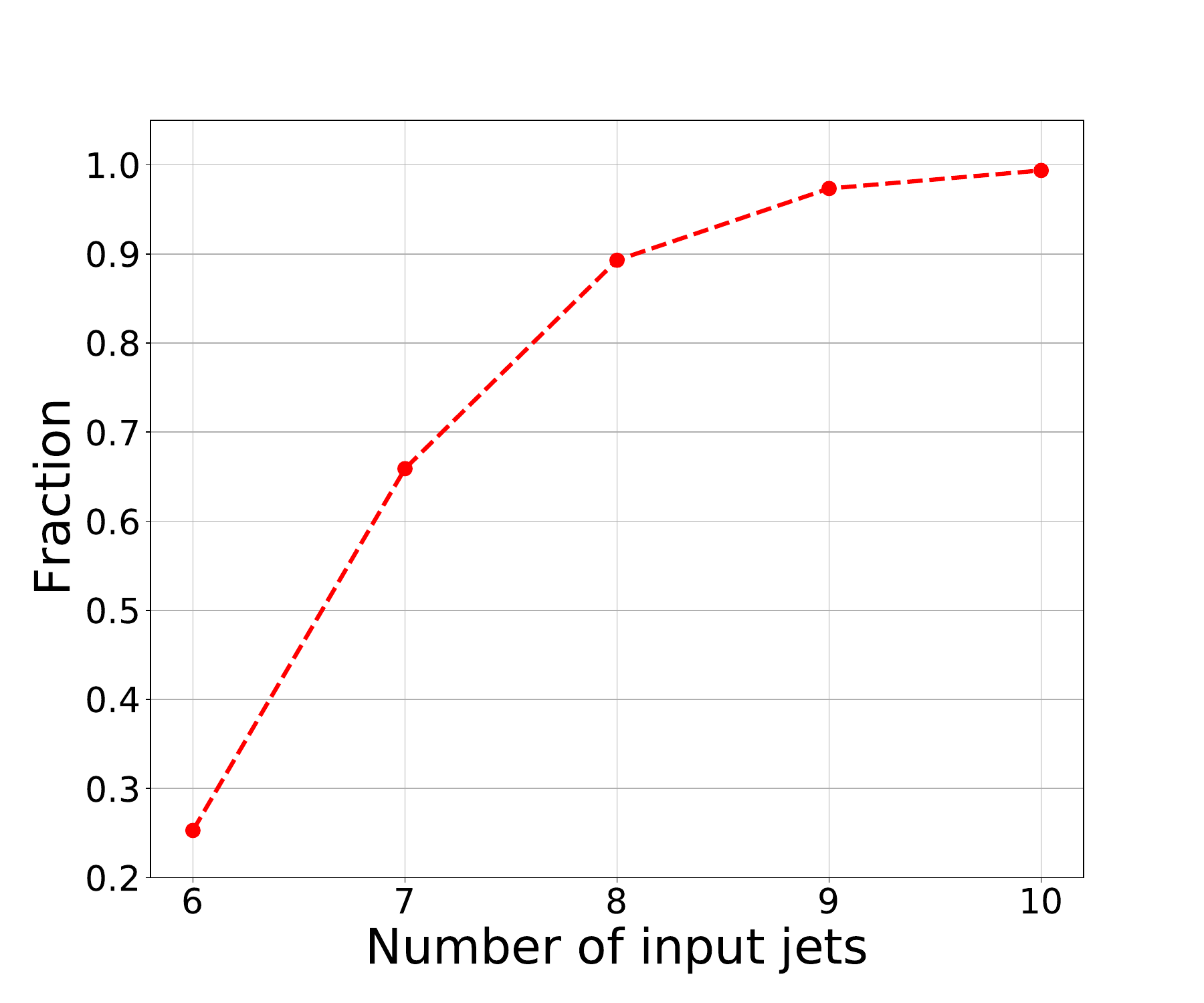}
    \caption{The jet multiplicity distribution for matched \ttbar\ (left). The fraction of matched \ttbar\ events, where all partons can be matched with most energetic $N$ jets without considering b-tag information (right).}
    \label{fig:jet-multiplicity}
\end{figure}

The jet multiplicity distributions for matched \ttbar\ events are shown in figure~\ref{fig:jet-multiplicity}.
The figure also shows the fraction of matched \ttbar\ events, where all partons can be matched with the most energetic $N$ jets without considering b-tag information.
These numbers can be interpreted as the naive upper limit of the performance of a jet-parton assignment model that takes the most energetic $N$ jets.


All jets in the event are used as input to the \textsc{SaJa} network. Jets are input into the network using high-level reconstructed variables. These are the jet \(p_{T},\ \eta,\ \phi\), \(\frac{p_{T}}{H_{T}}\), 
and whether the jet is b-tagged.

We also tested the impact of using additional jet shape variables.
Gluon-initiated jets should always be assigned to the ``other'' class, and these variables can be helpful to distinguish quark from gluon jets. 
We used eight variables, which are simple extensions of the three quark-gluon discrimination variables used by CMS~\cite{Cornelis:1754895} and shown to be effective in quark-gluon separation~\cite{lee2019quark}.
The variables are the jet energy sharing variables, major and minor axes of the jet in $\eta-\phi$ space, and the multiplicity of each particle flow type.
All features are scaled to the range from 0 to 1 using Min-Max scaling in order to stabilize the optimization process.

\begin{table}[tbp]
    \centering
    \begin{tabular}{|c|c|c|}
        \hline
        Hyperparameters      &  \textsc{SaJa} without Jet Shape & SAJA with Jet Shape \\
        \hline
        $d_{k}$              & 16                               & 32                  \\
        $N_{\textrm{head}}$  & 10                               & 10                  \\
        $d_{\textrm{FFN}}$   & 1024                             & 1024                \\
        $N_{\textrm{block}}$ & 6                                & 6                   \\
        \hline
    \end{tabular}
    \caption{\label{tab:hparams} Hyperparameters of \saja\ with and without jet shape.}
\end{table}

We used the Adam optimization algorithm~\cite{arXiv:1412.6980} to minimize the objective function given in eq.~\ref{eq:objective}. The learning rate started at 0.001 and was reduced by 2 when the loss measured on the validation set did not decrease over ten epochs to prevent a model from oscillating around local minima.
A batch of 512 randomly sampled events is used for each training step.
\textsc{PyTorch} v.1.4.0 is used for the implementation~\cite{NEURIPS2019_9015}.
Only the fully matched \ttbar\ events were used for training.
We used 310, 80, and 100 thousand events for training, validation, and testing, respectively.
The hyperparameters, listed in table~\ref{tab:hparams}, were optimized by hand as the search space is quite small. 


We also used a kinematic likelihood fit to do the jet-parton assignment to compare the performance of \textsc{SaJa}.
We used the \textsc{KLFitter} library to perform the kinematic likelihood fitting~\cite{erdmann2014likelihood}.
For the fit, the invariant mass of the $W$ boson is fixed to \(m_{W} = 80.4\) GeV, while the top quark invariant mass was floated in the fit.
We derived the transfer functions required by \textsc{KLFitter} from our top pair dataset.
The transfer function is modeled by a Crystal Ball function, whose parameters are parameterized by the energy of the matching parton with separate parameterizations for barrel and endcap jets.

As more jets are used, the number of permutations increases exponentially, and therefore, we could not use all the jets in an event to study the \textsc{KLFitter} performance.
We studied two different cases of jet inputs.
In the first case, the 6 most energetic jets in the event were used as inputs to the fit, which gives a total of 18 permutations.
In the second case, when more than 6 jets were in the event, we tried permutations of the 7 most energetic jets, giving an average of 126 permutations for the dataset.
The permutation with the lowest likelihood after fitting was taken as the jet-parton assignment for the event.
The value of the likelihood was also used as a goodness-of-fit measure.
When the best permutation has a likelihood lower than a threshold, the event is rejected.

\section{Results}
\label{sec:results}

\subsection{Performance}

Jet-parton assignments are used to reconstruct the signal topology by checking for topological validity, which also helps to reject background events.
Therefore, in this study, we measure not only the fraction of correct assignment for the matched \ttbar\ events but also the rejection rates for unmatched \ttbar\ and multijet events.

\begin{figure}[tbp]
    \centering
    \includegraphics[width=.48\textwidth]{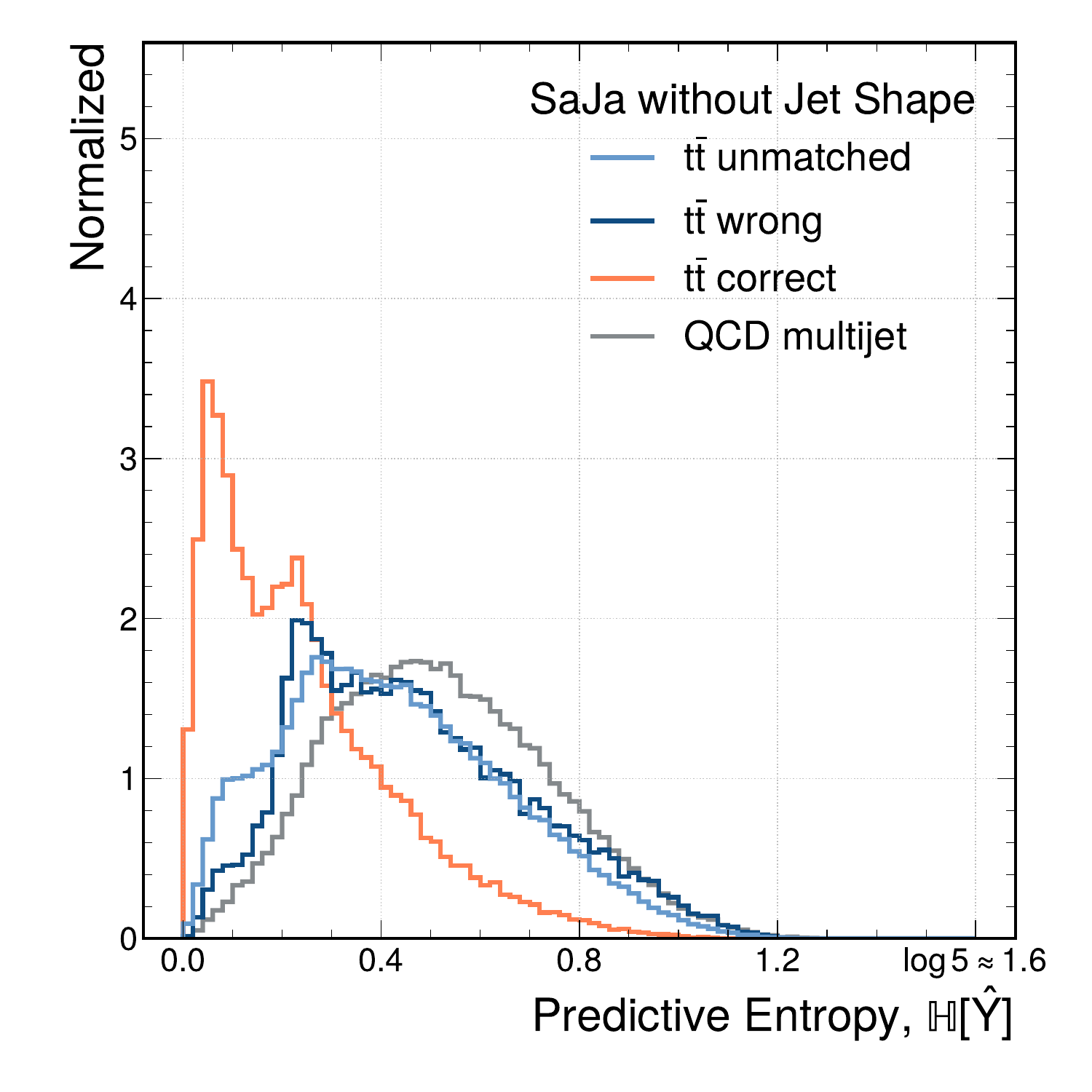}
    \includegraphics[width=.48\textwidth]{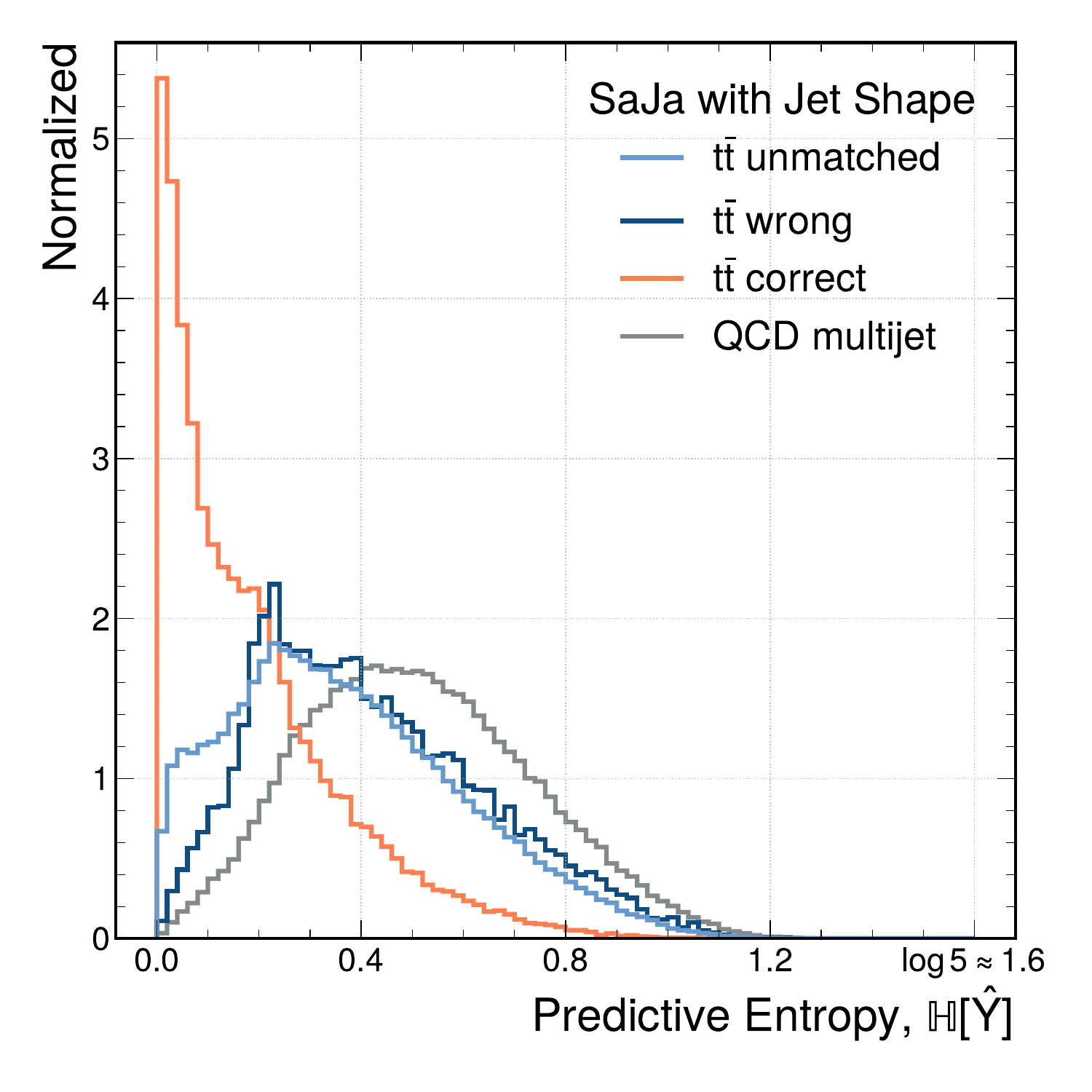}
    \caption{\label{fig:entropy} The prediction entropy distribution of \saja\ without jet shape (left) and \saja\ with jet shape (right). }
\end{figure}

Figure~\ref{fig:entropy} shows the distributions of predictive entropies for SaJa and SaJa with extra jet shape features. 
Correctly assigned \ttbar\ events have lower uncertainty than wrongly assigned \ttbar, unmatched \ttbar, or multijet events.
As expected, the predictive entropy can be used to effectively veto wrong assignments and background events.
For \saja\ without jet shape variables, the second peak near $\mathbb{H}[\hat{Y}]=0.2$ in the correct \ttbar\ assignment distribution is due to events where one or two of the $b$ jets are not $b$ tagged and there are no mistagged jets.
This undesired second peak is not observed in the case of \saja\ with jet shape variables, which implies the jet shape variables are giving \saja\ $b$ tag information.
Since \textsc{Delphes} only gives a binary classification for $b$ tagging, while in a real experiment the $b$ tag information is given as a probability, the performance of \saja\ without jet shape variables would be closer to our \saja\ plus jet shape variables case.
The multijet distribution is smoother and more Gaussian than the other distributions, peaking at a larger entropy point.
This means that the predictive entropy of \saja\ works well as an out-of-distribution detection method, even though it was only trained on matched $t\bar{t}$ events, without any additional training or uncertainty calibrations.
With an alternative training scheme incorporating the expected SM channels, this method has the potential as a model-agnostic approach for new physics search.

\begin{figure}[tbp]
    \centering
    \includegraphics[width=.48\textwidth]{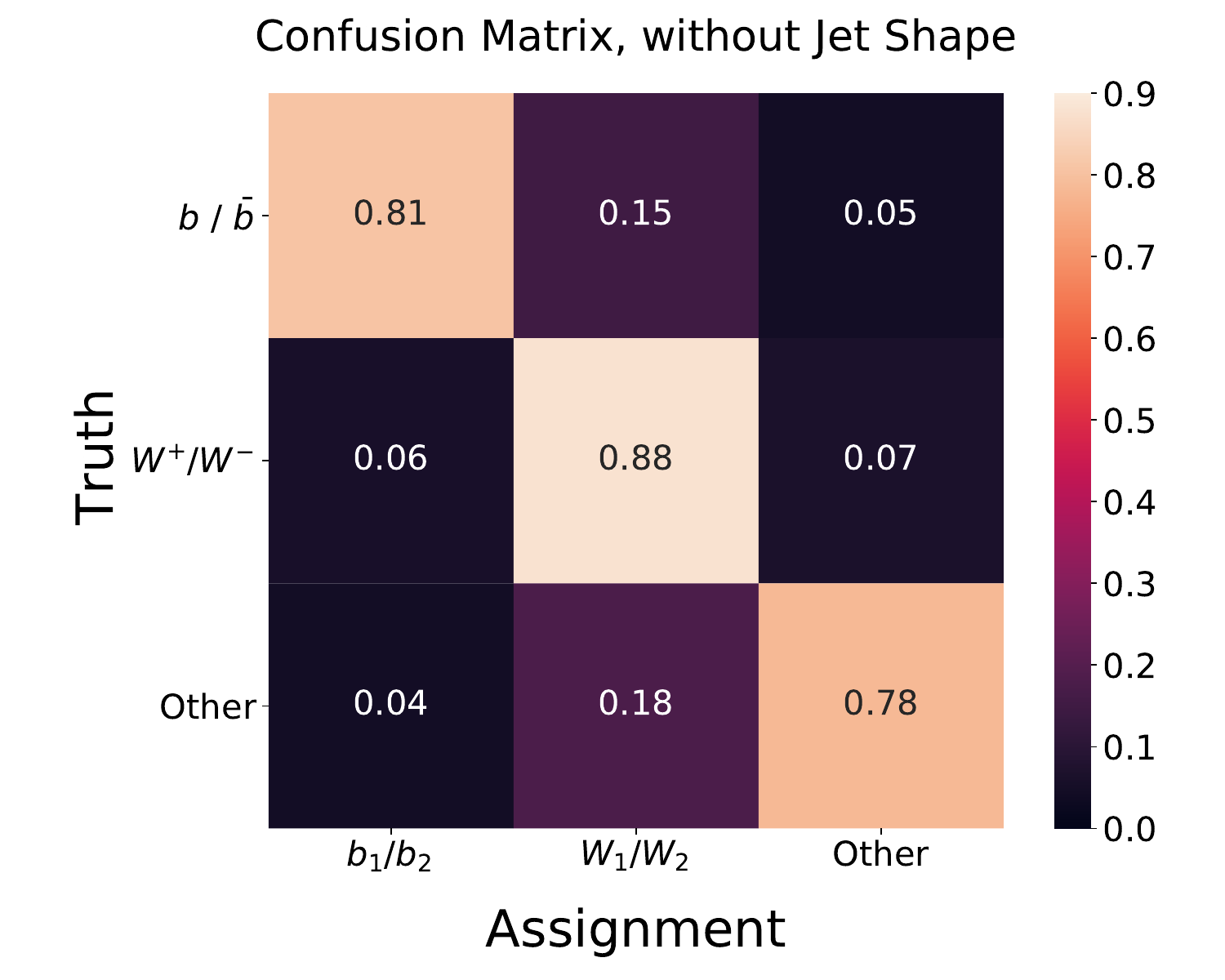}
    \includegraphics[width=.48\textwidth]{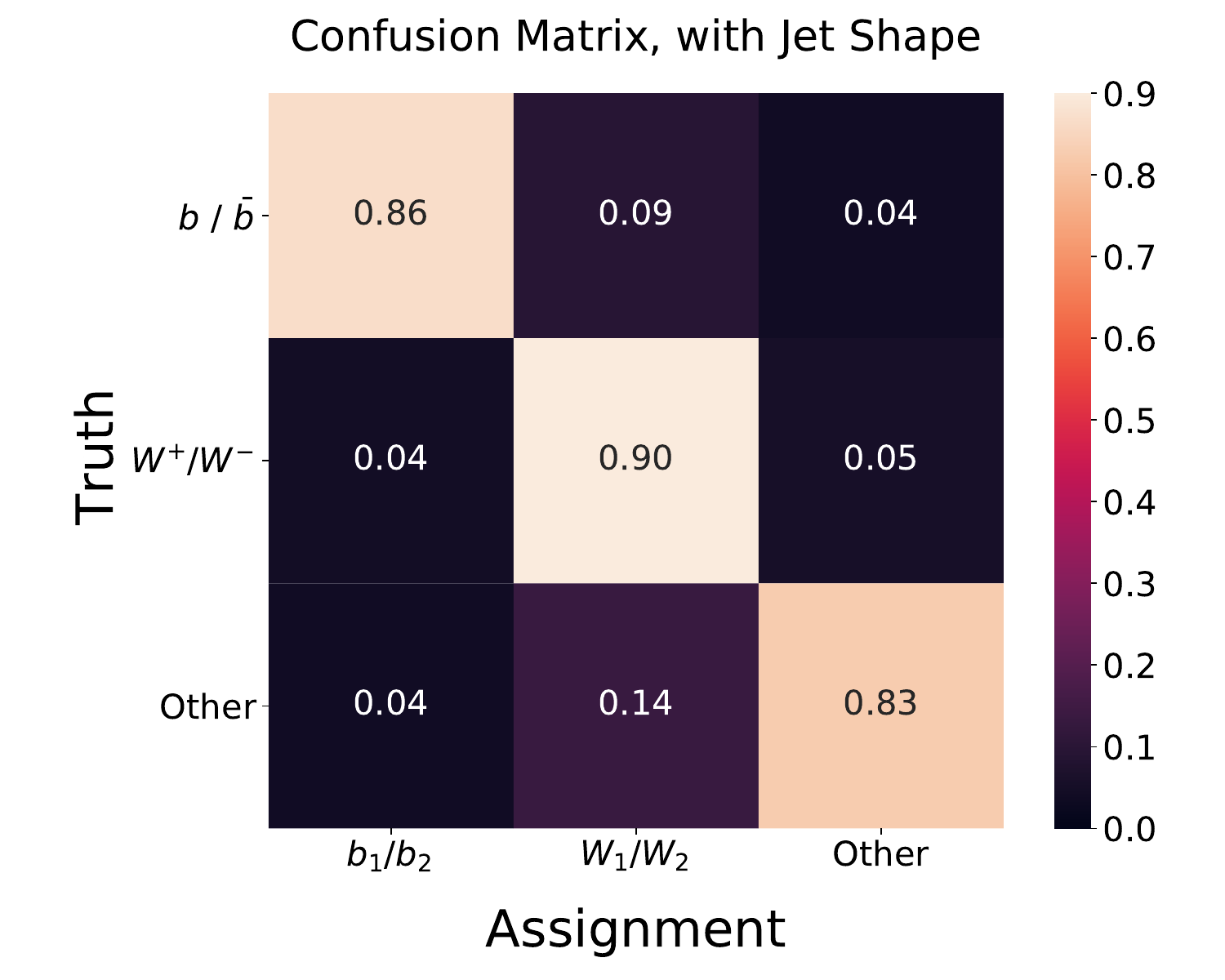}
    \caption{Jet-level normalized confusion matrices for \saja\ without jet shape (left) and with jet shape (right). The rows of the matrices correspond to the truth label given by the jet-parton matching and the columns do the prediction of \saja. }
    \label{fig:confusion-matrix}
\end{figure}

\begin{figure}[tbp]
    \centering
    \includegraphics[width=0.6\textwidth]{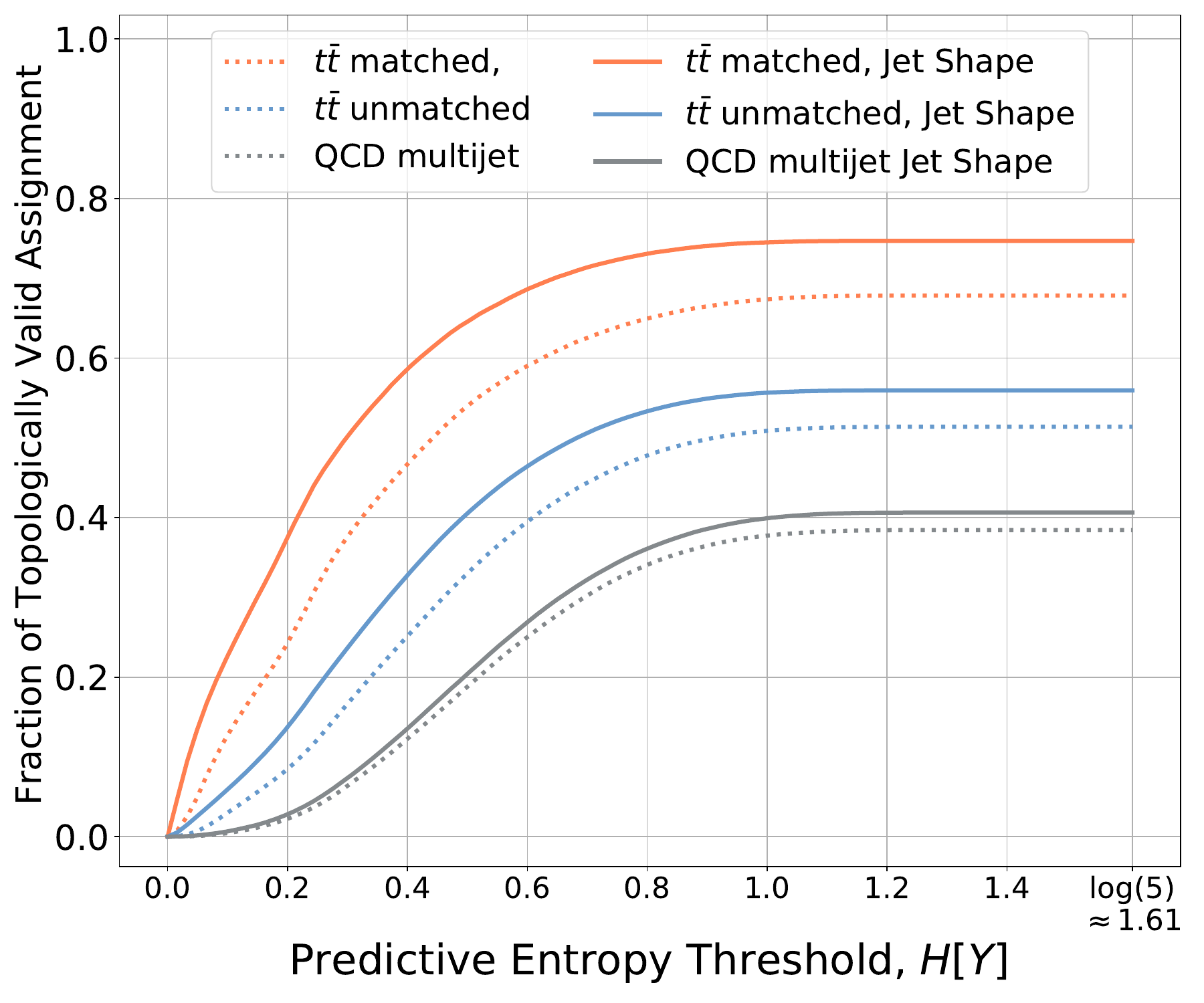}
    \caption{The fraction of topologically valid assignments passing the entropy threshold selection for \saja\ with jet shape information (solid) and without jet shape information (dotted) as a function of the threshold on matched \ttbar\ events (red), unmatched \ttbar\ events (blue), and multijet (black).}
    \label{fig:topo-valid}
\end{figure}

Figures~\ref{fig:confusion-matrix} shows jet-level normalized confusion matrices for \saja\ with and without additional jet shape variables.
Jet shape information increases the accuracy of predictions for b-jets and other jets by about 5\%.
This improvement is also shown in figure~\ref{fig:topo-valid}, the fraction of topologically valid assignments of \saja\ on the matched \ttbar, the unmatched \ttbar\, and multijet events.
For all cases, jet shape information increases the fraction of topologically valid events, however, the increase is larger for the matched events and the relatively small increase in multijet events can be easily rejected by adjusting the predictive entropy threshold.
The fraction of valid assignments gives a limit to the fraction of events we can correctly assign since invalid assignments are rejected from further analysis, as described in section~\ref{sec:objective}.

\begin{figure}[tbp]
    \centering 
    \includegraphics[width=.32\textwidth]{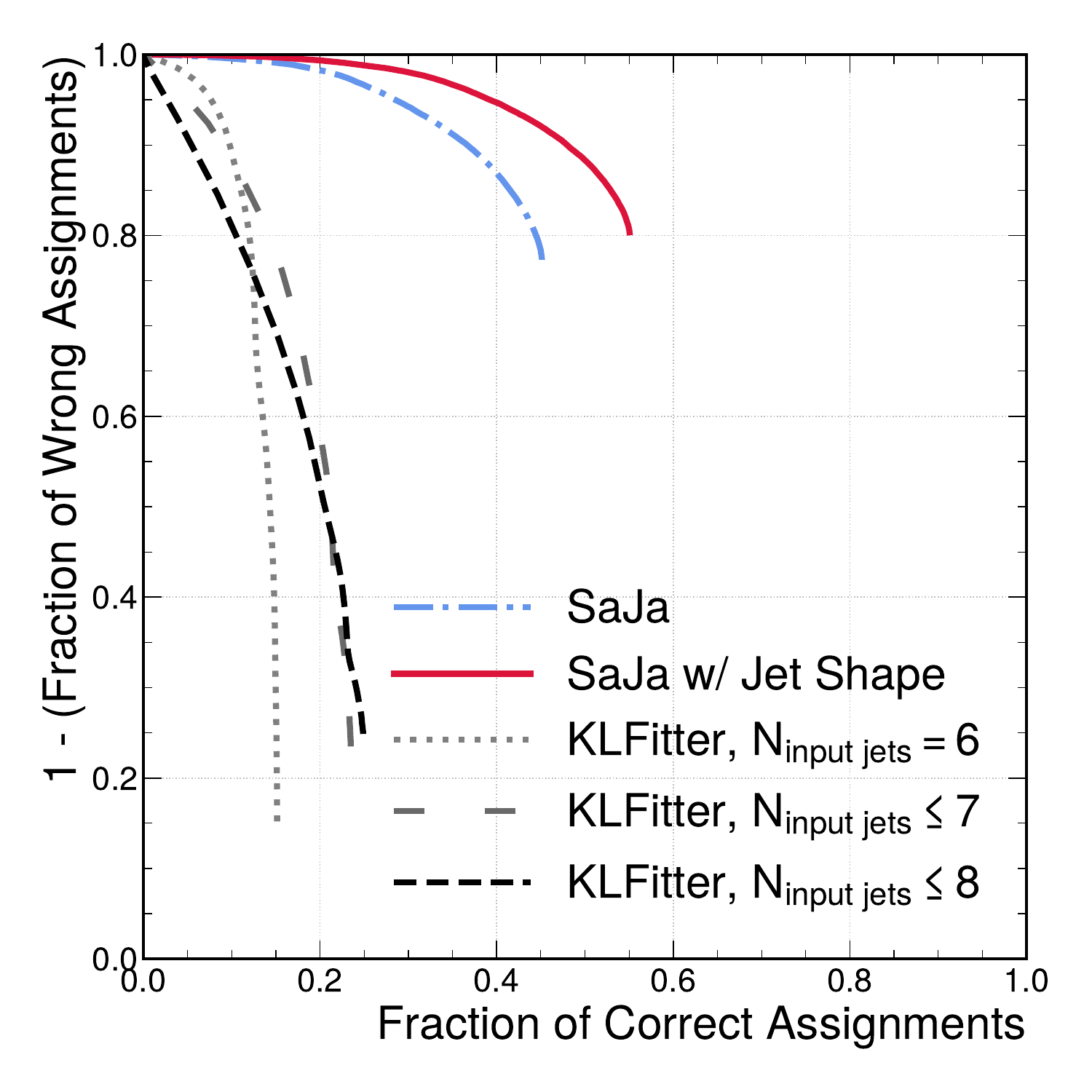}
    \includegraphics[width=.32\textwidth]{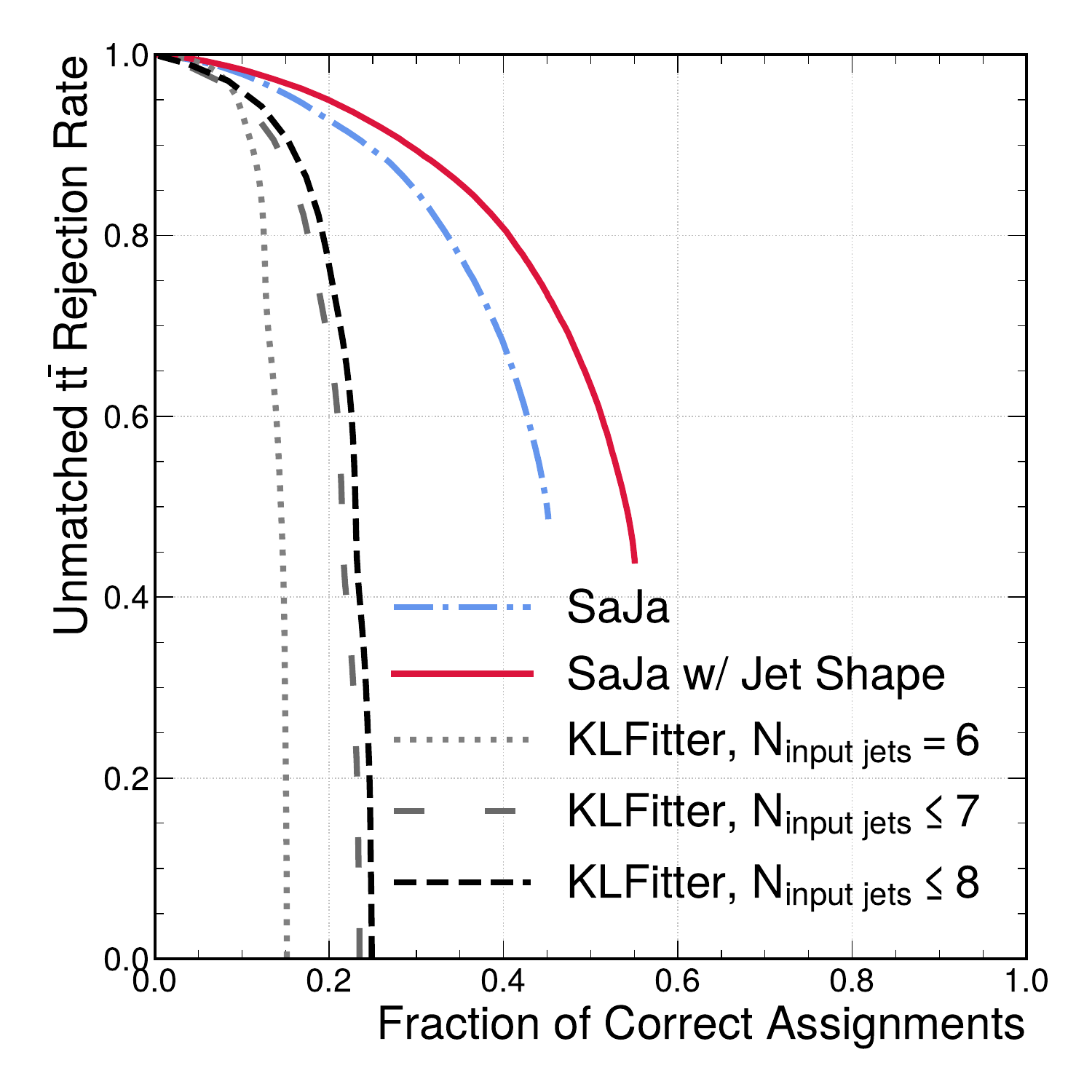}
    \includegraphics[width=.32\textwidth]{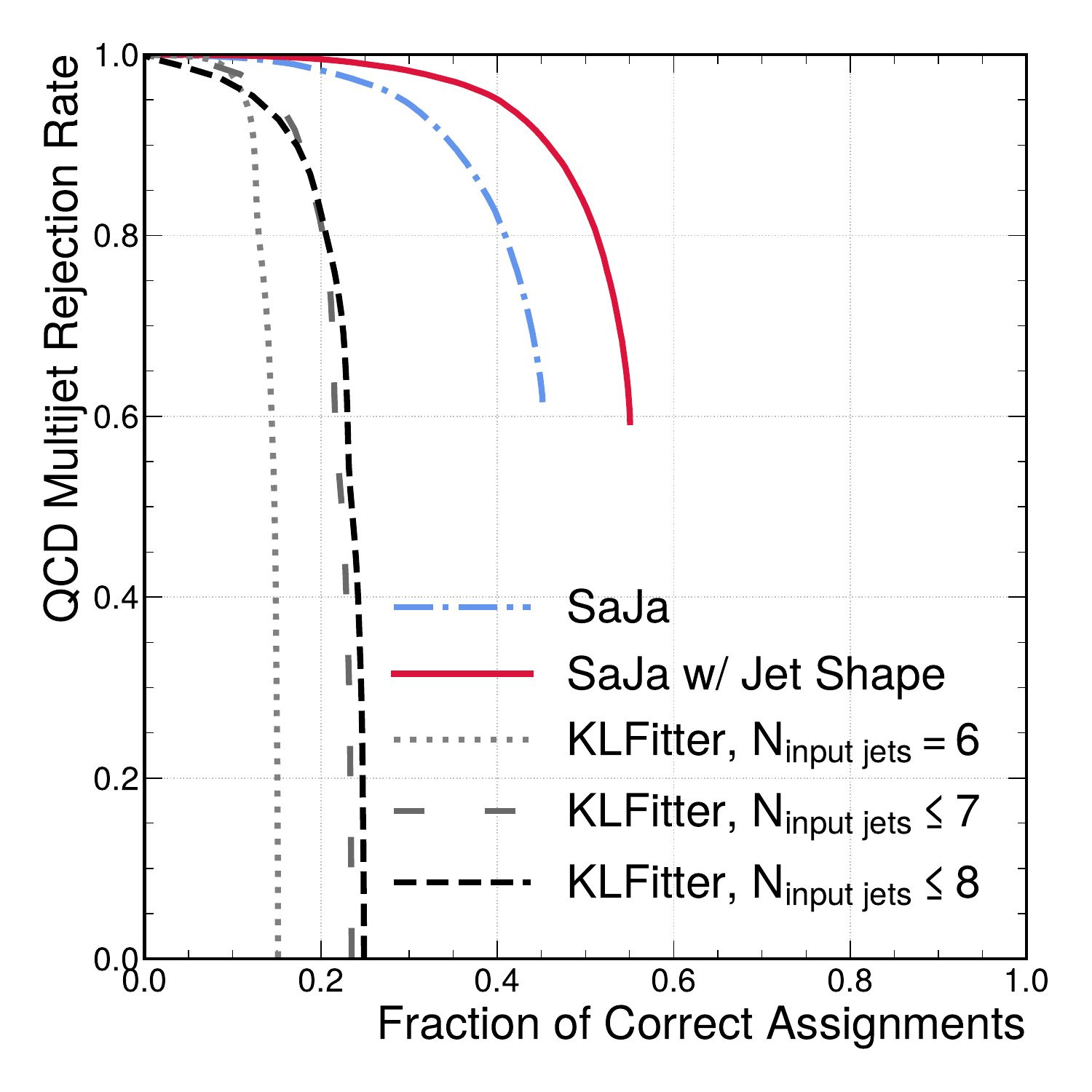}
    \caption{\label{fig:roc-model}
        The performance measurement curve. 1 - the fraction of wrong assignments for matched \ttbar\ (left), the unmatched \ttbar\ rejection rate (middle) and multijet rejection rate (right) as a function of the fraction of correct assignments for matched \ttbar.
    }
\end{figure}

The performance of \saja\ is compared to \klf\ using receiver operating characteristic curves, illustrated in Figure~\ref{fig:roc-model}.
The curves are constructed by the fraction of wrongly assigned matched \ttbar\ events and the background event rejection rates against the fraction of correct assignments for matched \ttbar\ events at various threshold points.
The predictive entropy and the negative-log likelihood serve as the thresholds for \saja\ and \klf, respectively.
The higher the curve is toward the upper right, the better the model performance, and the figures show that \saja\ exceeds the performance of \klf\ in all aspects.
Unlike general ROC curves, SaJa has cut short curves due to the rejection of topologically invalid assignments.

Figure~\ref{fig:roc-model} (left) also shows that \klf\ with more input jets performs worse as the negative log-likelihood threshold becomes tighter.
As \klf\ takes in more jets, the probability of a jet permutation involving a jet pair close to the W boson increases.
Since \klf\ evaluates jet permutations based on a few prior knowledge, like W boson mass, and cannot use the complex event topology, the wrong permutation can have a lower negative log-likelihood compared to the correct permutation.
This shows that a good jet-parton assignment algorithm needs to handle not only many jets efficiently but also the complex structure between them.

\begin{figure}[tbp]
    \centering
    \includegraphics[width=0.7\textwidth]{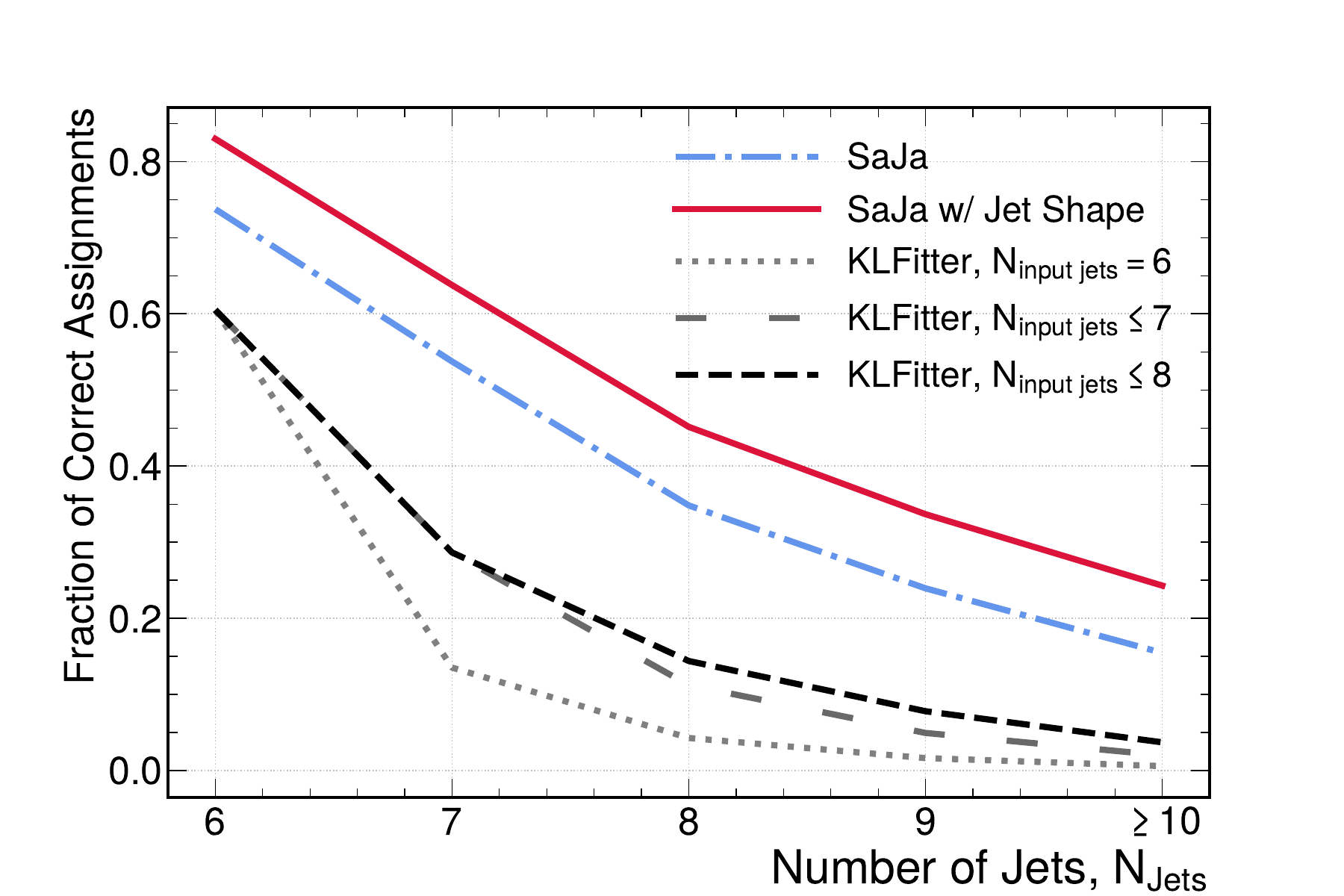}
    \caption{\label{fig:perf-vs-njets} The fraction of correct assignments versus the jet multiplicity of \saja\ without jet shape (blue), \saja\ with jet shape (red), \klf\ with up to 6 jets (black), and \klf\ with up to 7 jets (gray).}
\end{figure}

Figure~\ref{fig:perf-vs-njets} shows the fraction of correct assignments as a function of jet multiplicity. 
The performance difference between \saja\ and \klf\ are largest in $N_{\text{Jets}}=7$ and 8, which corresponds to the bulk of the \ttbar\ population that are matched, as shown in figure~\ref{fig:jet-multiplicity}.
\saja\ is able to better learn the underlying \ttbar\ event topology when given jet shapes.

\begin{figure}[tbp]
    \centering
    \includegraphics[width=0.48\textwidth]{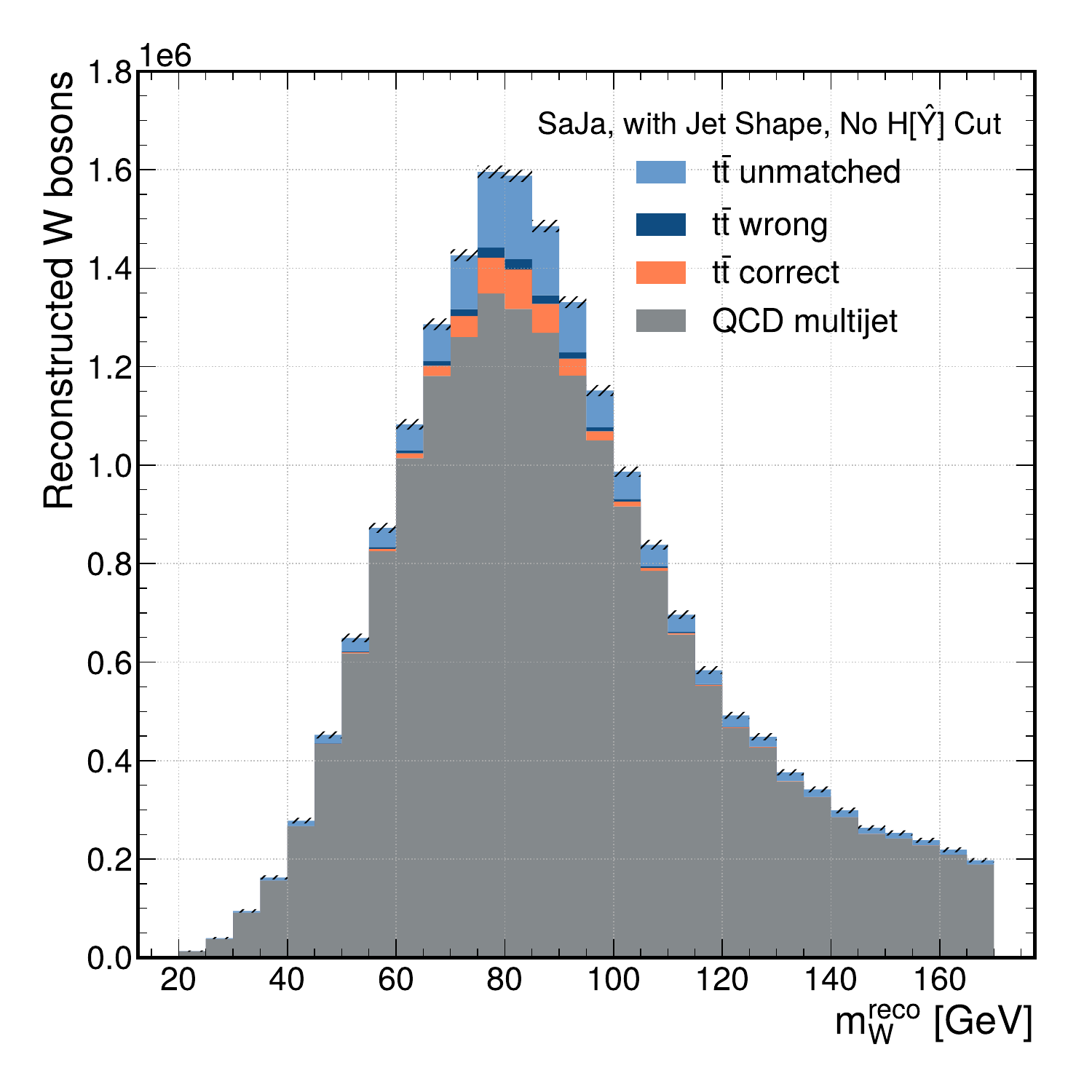}
    \includegraphics[width=0.48\textwidth]{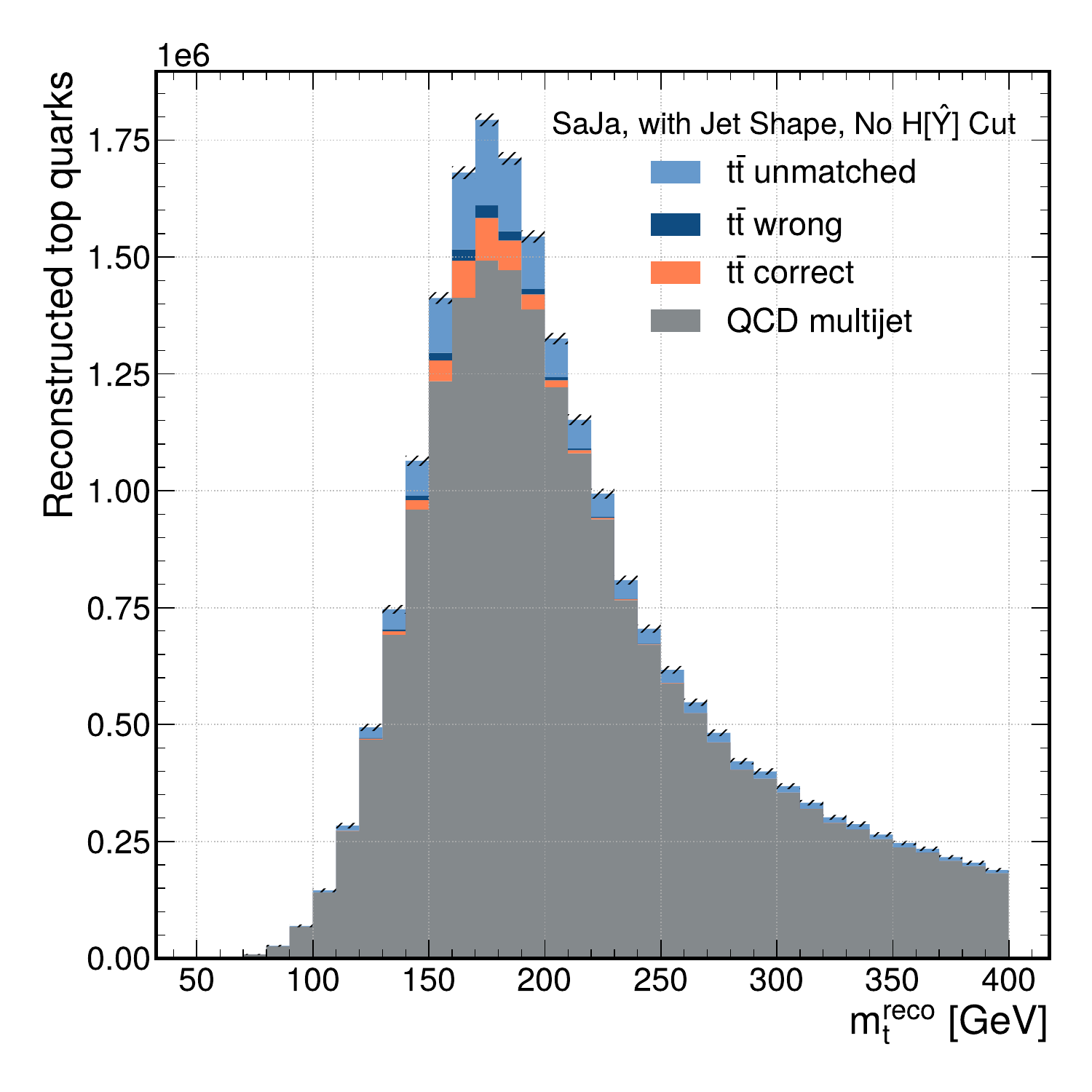}
    \hfill
    \includegraphics[width=0.48\textwidth]{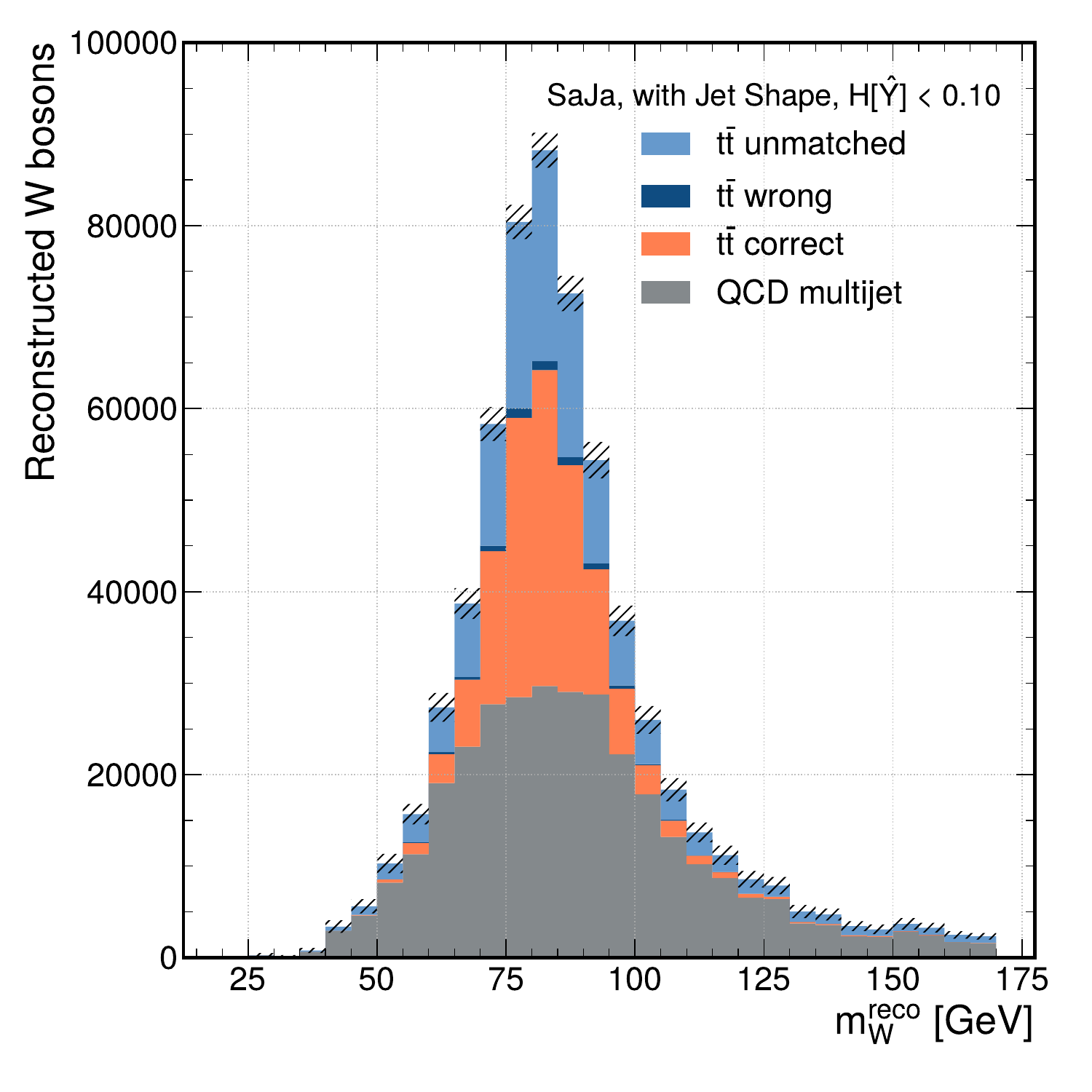}
    \includegraphics[width=0.48\textwidth]{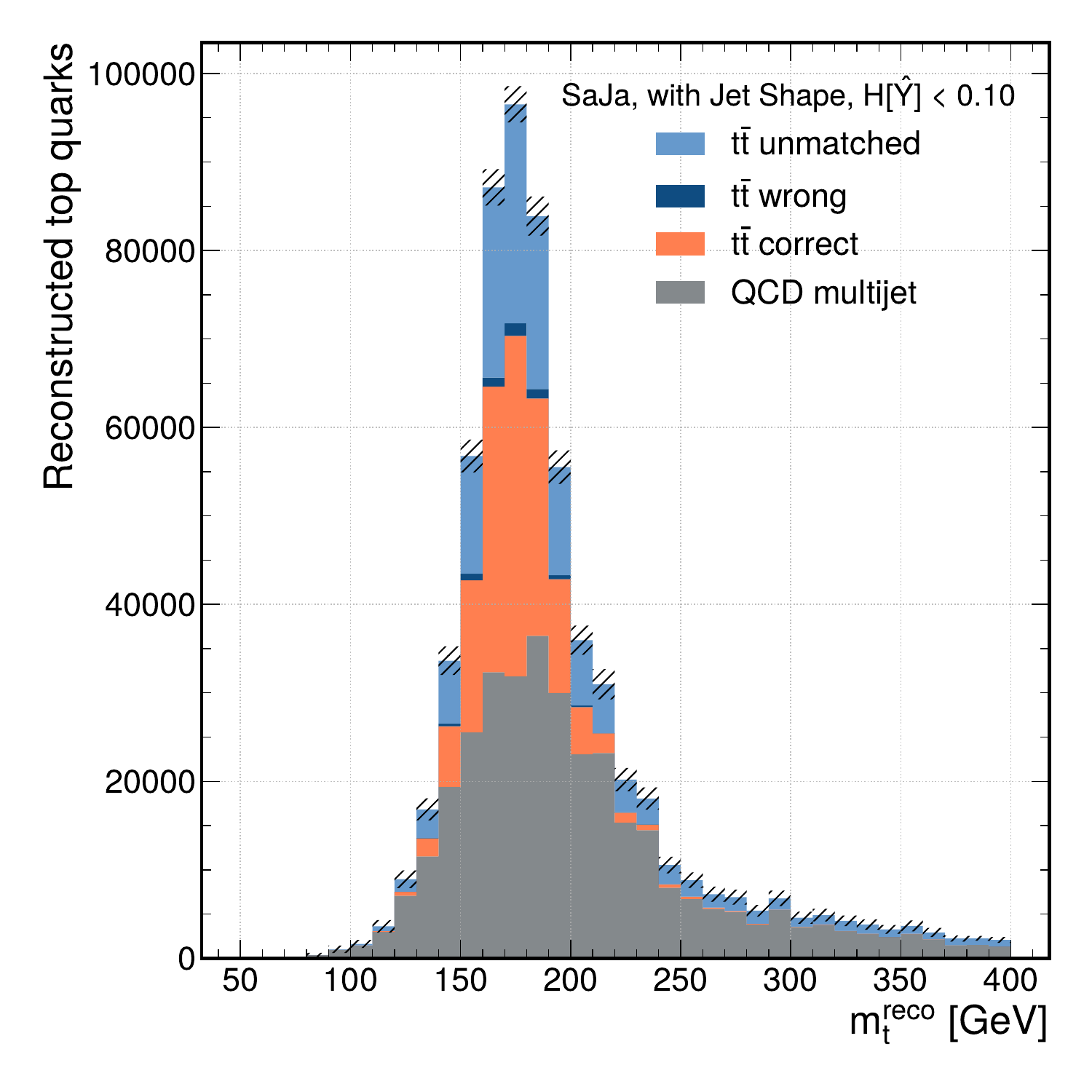}
    \caption{The reconstructed mass distribution of W boson (left) and top quark (right) both before (above) and after (below) the predictive entropy threshold selection. unmatched \ttbar\ (blue), wrongly assigned \ttbar\ (pink), correctly assigned \ttbar, and multijet (gray).}
    \label{fig:reco-mass}
\end{figure}
 
Figure~\ref{fig:reco-mass} shows the reconstructed W and top mass distributions obtained using \saja\ including jet shape information before and after a predictive entropy threshold of 0.074.
The distributions are normalized with the integrated luminosity of \(35.91~\textrm{fb}^{-1}\).
Clear peaks can be found in the W and top mass range. The QCD background is also accumulating in the top mass range, and this feature becomes more prominent after the selection with predictive entropy. This shows that the top mass is an important consideration for \saja\ to assign high certainty to an assignment.

\begin{figure}[tbp]
    \centering
    \includegraphics[width=0.7\textwidth]{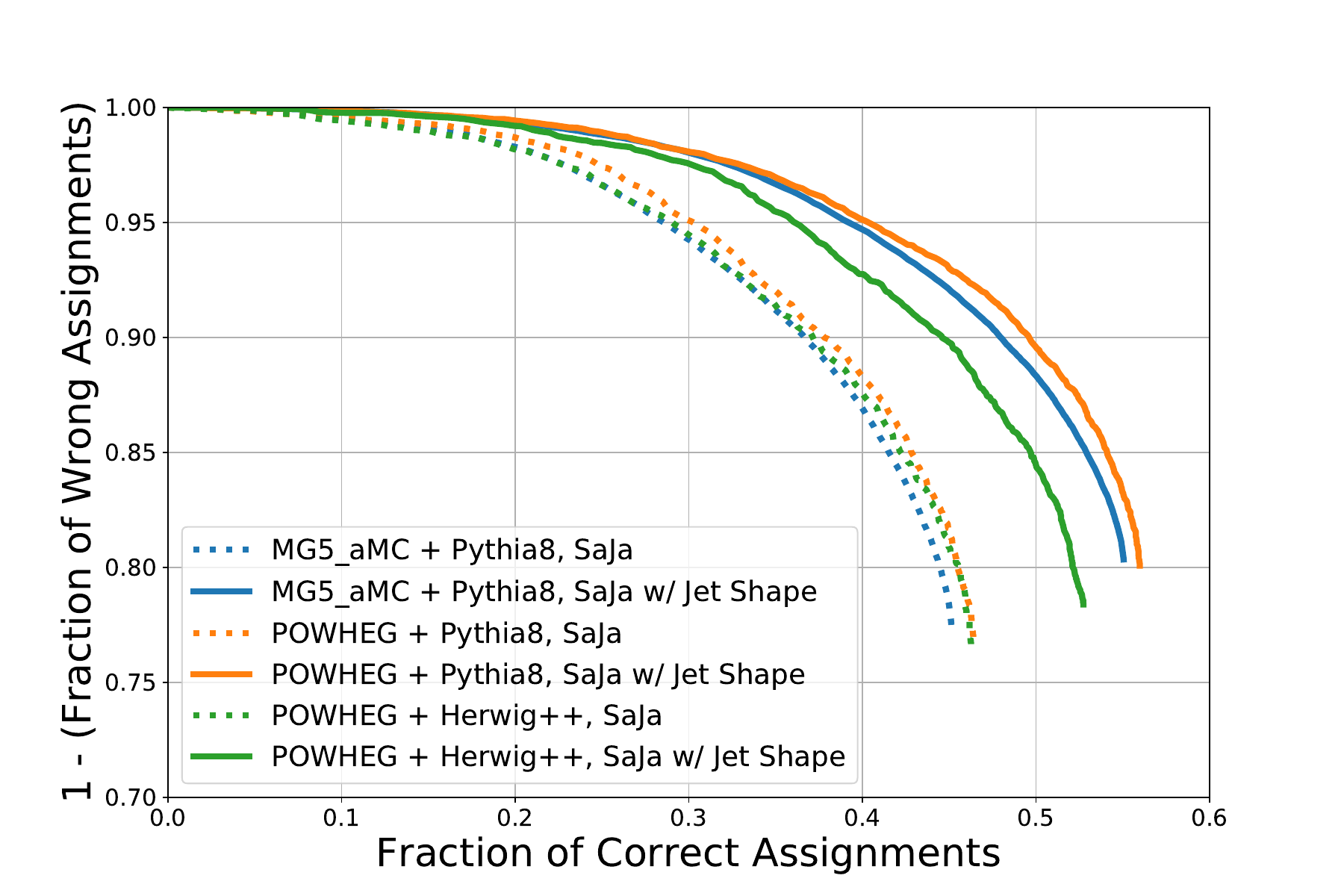}
    \caption{\label{fig:roc-generator} The fraction of wrong assignments as a function of correct assignments for matched \ttbar\ of \saja, which is trained on \textsc{MadGraph5\_aMC@NLO} interface to \textsc{Pythia8}.}
\end{figure}

The Monte Carlo dependency of \saja\ is shown in figure~\ref{fig:roc-generator}. It compares the ROC-like curves of figure~\ref{fig:roc-model} for the different generators: \madgraph\ interfaced to \pythia, \powheg\ interfaced to \pythia\ and \powheg\ interfaced to \herwig.
It can be seen from the figure that \saja\ without jet shape information is relatively insensitive to which event generator is used.
However, \saja\ with jet shape information shows a difference in performance, especially between \pythia\ and \herwig. This is to be expected due to their different parton shower models. 

\subsection{Model Interpretability}

\begin{figure}[tbp]
  \centering
  \includegraphics[width=0.7\textwidth]{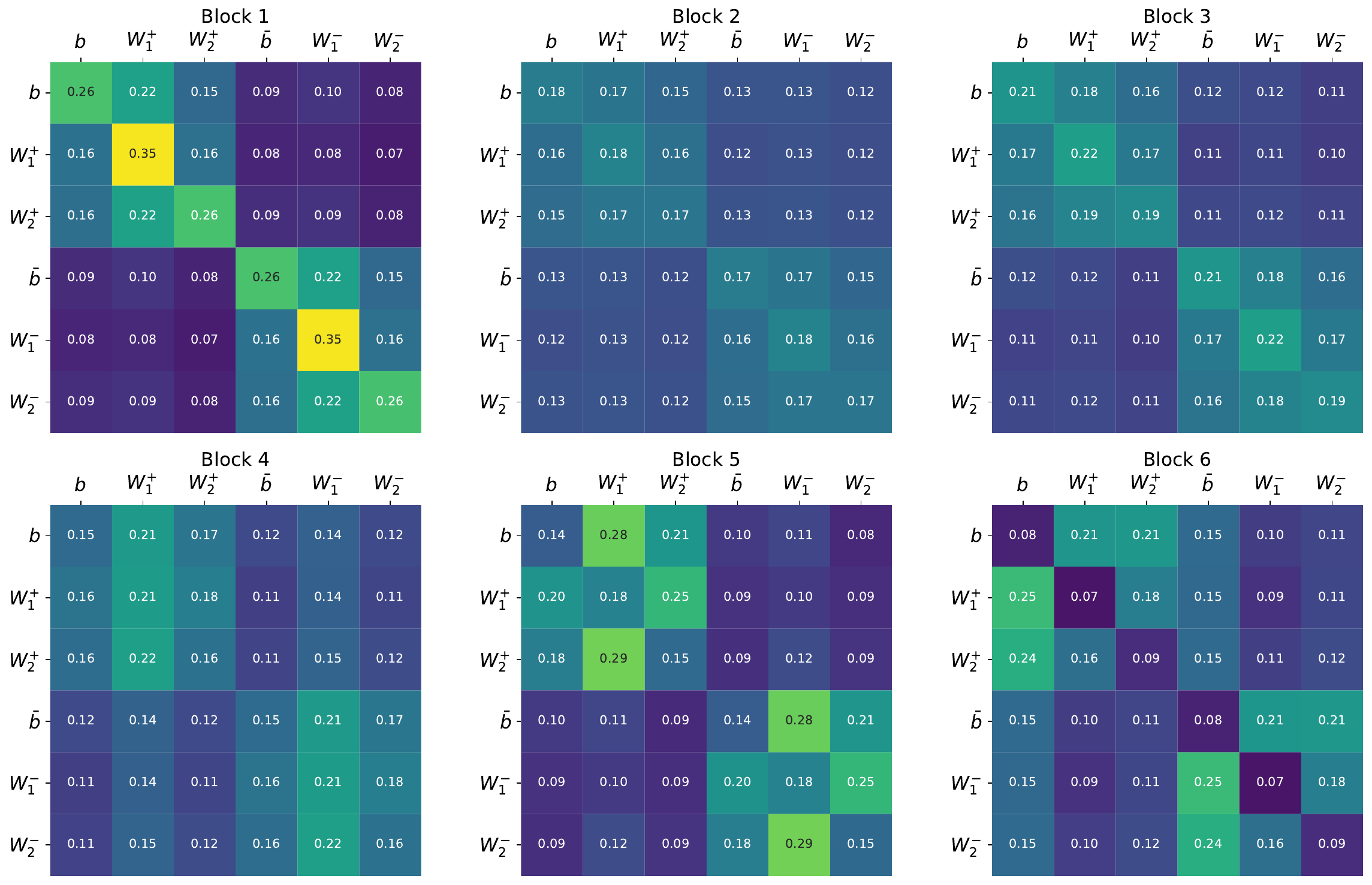}
  \hfill
  \caption{\label{fig:attention} The average of the attention weights block-by-block for the correct assignment in the matched \ttbar\, events. $b$ and $\bar{b}$ are bottom and anti-bottom quarks. $W_{1}^{+}$ is higher $p_{T}$ of the two quarks, the decay products of $W^{+}$ boson.}
\end{figure}

One of the benefits of the attention-based model is that the attention weights show the dependencies between the input elements \cite{arXiv:1409.0473,arXiv:1706.03762}.
Since the model is invariant under the permutation of the jets, it is possible to sort the jets using the jet-parton matching information without loss of generality, for display purposes.
Figure~\ref{fig:attention} shows the average of the attention weights block by block for \ttbar\ events where all jets are correctly assigned.
Attention weights act equivalently to the convolution kernel in convolutional neural networks, in that they weight the elements of the input sequence when creating the output.
That is, the weights indicate which keys are strongly influencing the output sequence for each query, both of which are derived from our input jets.
Rows correspond to queries and columns correspond to keys and so each row is normalized using the softmax function when creating the visualization.

In the first block, the diagonal elements have the highest values in each row and show strong attention between jets originating from the same top quark.
It can therefore be seen that the first block performs the jet representation learning.
The second to fourth blocks show relatively uniform attention weights and seem to perform the event-level representation learning.
The fifth and sixth blocks show strong attention to jets originating from the same top quark again.
Also, in the last sixth block, the diagonal component is very small.
So, the fifth and sixth blocks extract the information required for the jet-parton assignment.

This model's interpretability, enhanced using Monte Carlo information, can be used for the transfer learning of other event properties.
For example, we can remove the top two layers and then use \saja\ as a pre-trained backbone to speed up the training of full-event classification or for a regression output targeting the kinematics of the reconstructed top.


 \subsection{Latency}
 \begin{figure}[tbp]
  \centering
  \includegraphics[width=0.7\textwidth]{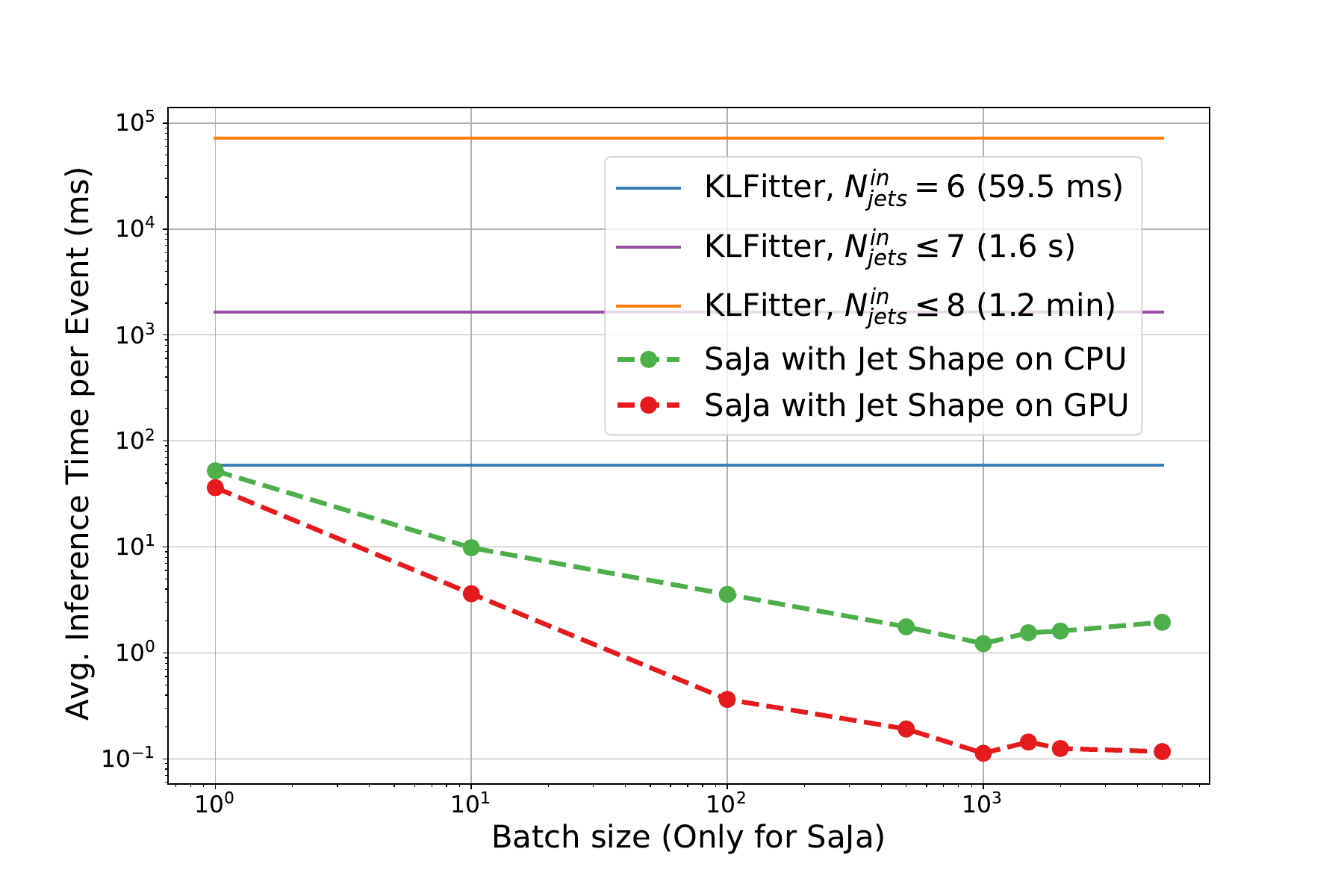}
  \hfill
  \caption{\label{fig:latency} The average inference time per event. For \saja\, this is shown as a function of batch size. For \klf\ the events must be processed serially, so there is only a single item shown, but it is extended across the plot for comparison with the \saja\ results.}
 \end{figure}
 Figure \ref{fig:latency} shows the average inference time per event, comparing \saja\ and \klf.
 We used a NVIDIA V100 GPU, and a Intel Xeon E5-2630 v2 @ 2.60 GHz CPU.
 The \saja\ network can be processed in batches, allowing for a large speedup, especially on the GPU, while the \klf\, fits must be performed one at a time, and so have fixed times.
 The graph shows that even for a batch size of one, the zero-permutation \saja\ network is outperforming \klf, while a two-order-of-magnitude speed up in inference time is possible by fitting in large batches.
 When \saja\ with jet shape takes a batch containing 1000 events of up to 12 jets, the maximum GPU memory allocated for the tensor, which is the model's parameters and all the inputs, is about 181 MB.

\section{Conclusions}

In this paper, we introduced the \saja\ network for jet-parton assignment in high-energy physics events.

\saja\ performed better than the traditional kinematic likelihood methods, as implemented in \klf.
The \saja\ network shows great potential for future use, as it could be easily adapted to other topologies of arbitrary complexity, such as fully-hadronic \(t\bar{t}H\), for which traditional machine learning techniques have been computationally expensive. 

\section*{Acknowledgements}
This article was supported by the computing resources of the Global Science Experimental Data Hub Center at the Korea Institute of Science and Technology Information and the University of Seoul.
JL and SY are supported by the National Research Foundation of Korea (NRF) grant funded by the Ministry of Science and ICT (MSIT) (2019R1C1C1009200).
IW is supported by the National Research Foundation of Korea (NRF) grant funded by the Ministry of Science and ICT (MSIT) (2018R1C1B6005826).
IP is supported by the Basic Science Research Program through the NRF funded by the Ministry of Education (2018R1A6A1A06024977).




\bibliography{references.bib}

\nolinenumbers

\end{document}